\documentclass[a4paper,fleqn,usenatbib]{mnras}

\usepackage{newtxtext,newtxmath}

\usepackage[T1]{fontenc}
\usepackage{newtxtext,newtxmath}
\usepackage{ae,aecompl}
\usepackage{placeins}
\usepackage{graphicx}	
\usepackage{amsmath}	
\usepackage{amssymb}	
\usepackage{hyperref}
\usepackage{gensymb}

\title[Wobble damping]{Simulations of wobble damping in viscoelastic rotators}
\author[Quillen et al.]{
Alice C. Quillen$^{1}$\thanks{E-mail: alice.quillen@rochester.edu},
Katelyn J. Wagner$^{1,2}$ and 
Paul S\'anchez$^3$
\\
$^{1}$Department of Physics and Astronomy, University of Rochester, Rochester, NY 14627, USA\\
$^{2}$Department of Computer Science, Mathematics and Physics, Roberts Wesleyan, Rochester, NY 14624, USA\\
$^{3}$Colorado Center for Astrodynamics Research, The University of Colorado Boulder, UCB 431, Boulder, CO 80309-0431, United States\\
}

\hypersetup{draft}

\begin{document}
\maketitle
\begin{abstract}
Using a damped mass-spring model, we simulate wobble of spinning homogeneous  viscoelastic ellipsoids undergoing non-principal axis rotation. Energy damping rates are measured for oblate and prolate bodies with different spin rates, spin states, viscoelastic relaxation timescales, axis ratios, and strengths. Analytical models using a quality factor by Breiter et al. (2012) and for the Maxwell rheology by Frouard \& Efroimsky (2018) match our numerical measurements of the energy dissipation rate after we modify their predictions for the numerically simulated Kelvin-Voigt rheology. Simulations of nearly spherical but wobbling bodies with hard and soft cores show that the energy dissipation rate is more sensitive to the material properties in the core than near the surface. 


\end{abstract}


\section{Introduction}

Increasingly large multiple epoch photometric surveys, giving light curves with many data points,
are adding to the population of asteroids and comets
with rotational measurements (e.g., \citealt{warner09,masiero09,waszczak15,vaduvescu17,chang17}).
The recent discovery of Interstellar object 1I/2017 U1  'Oumuamua with extreme axis ratios and 
non-principal axis  rotation \citep{meech17,drahus18,fraser18,belton18} motivates
studying wobble damping in elongated bodies.
About 1 of 100 asteroids in the Asteroid Light Curve Database (LCDB) \citep{warner09} are
 identified as non-principal axis  rotators \citep{pravec14}. 
 
Development of a viscoelastic mass spring model for spinning gravitating bodies \citep{quillen16_crust,frouard16,quillen16_haumea,quillen17_pluto,quillen19_bennu,quillen19_moon}
allows us to numerically study spin evolution with a code that 
directly relates simulated material strength and dissipation rate  to variations in the spin state. 
Our numerical simulations are complimentary to analytic computations of the wobbling 
or precession damping rate
\citep{burns73,sharma05,efroimsky00,breiter12,frouard18}.
Moreover, the simulations are flexible as they can be used
to model complex body shapes (e.g., \citealt{quillen19_bennu}),
bodies with inhomogeneous internal composition (e.g., \citealt{quillen16_haumea})
and the distribution of internally dissipated heat (e.g., \citealt{quillen19_moon}).

At a given total rotational angular momentum, 
the most stable rotational state of a rigid object is that with the least rotational kinetic energy. 
This corresponds to the rotation about the axis of maximum moment of inertia 
or about its shortest principal body axis.
A body that is initially rotating about a principal axis can be rotationally excited due to out-gassing 
(if a comet, \citealt{marsden73}), an impact (e.g., \citealt{paolicchi02}),  internal shape changes \citep{sanchez18},
 or  applied external torques  such as a gravitational torque during a close approach
with another massive object,  (e.g., \citealt{scheeres00}).
If the rotation axis differs from a principal body axis, the spin state is described as 
a non-principal axis ({\bf NPA}) rotation state.   
NPA rotational states are also known in the literature as complex rotational states 
or objects that are tumbling or wobbling. 

Due to cyclic variations in stresses and strains, an object in an NPA state
slowly looses kinetic energy \citep{prendergast58}.
For an object to be in an NPA state, the damping timescale due to dissipation or mechanical friction 
must be larger than either the time elapsed since the last spin excitation event 
(e.g., such as  an asteroid impact) or a time scale associated with rotational excitation by 
another process, such as out-gassing. 
Some well studied small bodies are in NPA states.  
Examples are Comet 1P/Halley (e.g., \citealt{belton91,samara91}) 
asteroid 4179 Toutatis \citep{hudson95}, asteroid 99942 Apophis \citep{pravec14} and 
Interstellar object 1I/2017 U1 'Oumuamua \citep{drahus18,fraser18,belton18}.
 
An estimate for the damping timescale for a wobbling spinning body is
\begin{equation}
\tau \sim \frac{\mu Q}{\rho K r_n^2  \Omega^3}, \label{eqn:burns}
\end{equation}
\citep{burns73}
where $r_n$ is an average radius,  $\rho$ is a bulk density, $\mu$ is a shear modulus, $Q$ is an energy dissipation factor,
also called a quality factor, 
and $\Omega$ the spin rate.  The scaling factor $K$ takes into account
body shape and initial spin state.    
The product $\mu Q \sim 10^{11}$ Pa 
is representative of small solar system bodies \citep{harris94,pravec14}.
In the body's reference frame of a homogeneous body with rotational symmetry, such as an oblate or prolate ellipsoid, 
the spin axis precesses with a single frequency about the
body's axis of symmetry.  This precession frequency gives
the period for stress-strain cycling within the asteroid. 
The energy dissipation parameter $Q$ (also called the quality factor) describes the fraction 
of energy lost per spin or spin precession period \citep{efroimsky00,sharma05,breiter12}.  
More generally, the energy dissipation rate in a wobbling body is a function of the current spin rate and state,
the ratios between the body's three moments of inertia  and the rheology.  
If a body has three different moments of inertia (e.g., is a triaxial ellipsoid), 
the spin vector in the body frame expanded in Fourier series contains multiple frequency components.
The frequency dependence of the energy dissipation rate affects the wobble damping timescale
so triaxial bodies are more difficult to model than prolate or oblate ellipsoids of revolution 
\citep{efroimsky00,sharma05,breiter12,frouard18}.

The paucity of asteroids above a spin limit of 2.2 hours and distribution of body axis ratios at each spin period, requiring shear strength but little tensile strength, 
has lead to a granular aggregate or rubble pile interpretation of asteroid composition \citep{walsh18}. 
Wobble damping computations assume that spin associated accelerations cause
 small  variations in strain and that the body is  `prestressed'  by its own self-gravity \citep{sharma05}.
The prestressing means that cohesive forces between grains are not necessarily required for elastic behavior.
In this context the elastic shear modulus, $\mu$, characterizes variations in the repulsive elastic contacts
 between grains, pebbles and boulders
about the mean stress field.  There is considerable uncertainty about the viscoelastic behavior of asteroid material.
A strong pressure pulse in asteroid material could be rapidly attenuated, as they are in laboratory granular materials \citep{odonovan16}
and giving a low $Q$ for stress strain cycling.  Alternatively, 
seismic attenuation rates are long in lunar regolith \citep{dainty74,toksoz74,nakamura76}  compared to typically seen
on the Earth and if asteroid material acts like lunar regolith it might have a large $Q$ of a few thousand.
The P and S-wave speeds in a loose granular material could depend upon pressure and porosity 
\citep{hostler05}.

Going beyond a $Q$ quality factor description of energy dissipation, \citet{frouard18} have
calculated the wobbling damping, or nutation damping timescale for homogeneous oblate ellipsoids that obey a 
Maxwell viscoelastic rheology.
They found that the energy dissipation rate and associated tumbling damping timescale is primarily a function of
viscosity $\eta$ with  $1/\eta$ 
replacing the ratio of a spin frequency to the  shear modulus times the quality factor, $\mu Q$,  
in the predicted dissipation rate.
Their theory can be extended for any linear rheology, where there is a linear relation between stress
and strain.

Instead of analytically computing the dissipation rate
(as did \citealt{efroimsky00,sharma05,breiter12,frouard18}), we directly simulate wobble damping
in an extended body using an N-body and mass-spring model.
The mass-spring model is an N-body simulation technique of low computational complexity \citep{quillen16_crust,frouard16}.
As spring forces are applied between pairs of point particles, the simulation 
 accurately conserves angular momentum.  Since the strain of a spring
is computed from the distance between a pair of point particles, the simulations
  can measure extremely small levels of strain.  Thus mass-spring N-body models are an attractive
method for simulating rotational dynamics.
Because our mass-spring model adds both spring damping forces and spring elastic forces on pairs of particles,
our model approximates a Kelvin-Voigt viscoelastic rheology \citep{frouard16} rather than the
Maxwell rheology studied by \citet{frouard18}.  
Dissipation rates as a function of frequency can be compared between the two viscoelastic models using the 
complex compliances for each model, relating stress to strain as a function of frequency (e.g., see appendix D by \citealt{frouard18}).

The discovery of active asteroids P/2013 R3 and P/2013 P5 \citep{jewitt13,jewitt14}
motivates the study of the disruption of inhomogeneous self-gravitating aggregates. 
Following Yarkovsky-O'Keefe-Radzievskii-Paddack effect (YORP) induced spin increase, 
the mode of deformation or disruption depends on 
the strength, density and cohesion of an internal core, compared to that of a shell \citep{hirabayashi14,hirabayashi15,hirabayashi15b,scheeres15,sanchez18}.
Asteroids may host a surface layer of more highly dissipative regolith over a harder, less dissipative 
but fractured core \citep{nimmo17}.
In our previous study of tidal spin down of triaxial ellipsoids undergoing principal axis rotation, 
we found that the energy distribution rate was sensitive to small volumes of low strength material \citep{quillen16_haumea}.
Internal variations in strength and viscosity should also affect the wobble damping rates.
Our code is not restricted to homogeneous material properties or ellipsoidal shapes so we can numerically explore
the sensitivity of wobble damping to core strength.


In this paper we  describe our simulations in section \ref{sec:sims}.
We discuss numerical measurements of energy dissipation rates for homogeneous ellipsoids undergoing NPA rotation. 
In section \ref{sec:ob_pro},
we modify the predictions by \citet{frouard18,breiter12} for our simulated viscoelastic rheology and
compare them to our numerically measured dissipation rates.
In section \ref{sec:core}, we discuss numerical experiments on inhomogeneous bodies. 
A summary and discussion follows in section \ref{sec:sum}.
Nomenclature is summarized in Table \ref{tab:nomen}.

\section{Elastic Body Simulations}
\label{sec:sims}

\subsection{Mass-spring models}
\label{sec:sim_description}

To simulate energy dissipation and spin evolution of a non-spherical body in a non-principal
axis rotation state,
we use a mass-spring model \citep{quillen16_crust,frouard16,quillen16_haumea,quillen17_pluto},
that is built on the modular N-body code rebound  \citep{rebound}.
A viscoelastic solid is approximated as a collection of mass nodes that are
connected by a network of springs.
Springs between mass nodes are damped and
 the spring network approximates the behavior of a Kelvin-Voigt viscoelastic
solid with Poisson ratio of 1/4 \citep{kot15}.
When a large number of particles is used to resolve the spinning body, and with many 
springs in the spring-network, the mass-spring model behaves
like an isotropic continuum elastic solid \citep{kot15}, including its ability to transmit seismic waves
and vibrate in normal modes \citep{quillen19_bennu}. 

The mass particles or nodes in the resolved spinning body are subjected to three types of forces:
the gravitational forces acting on every pair of particles in the body, and the elastic and damping spring forces
acting only between sufficiently close particle pairs.
As all forces are applied equal and oppositely to pairs of point particles and along the direction
of the vector connecting the pair, momentum and angular momentum conservation are assured.
Springs have a spring constant $k_s$ and a damping rate parameter $\gamma_s$.
The number density of springs, spring constants and spring lengths set the shear modulus, $\mu$, whereas
the spring damping rate, $\gamma_s$, sets the shear viscosity, $\eta$,
and viscoelastic relaxation time, $\tau_{\rm relax} = \eta/\mu$.
For a Poisson ratio of 1/4, the Young's modulus $E = 2(1+ \nu) \mu = 2.5 \mu$.
The equation we used to calculate Young's $E$ from the spring constants of the springs in our code is equation 29
by \citet{frouard16} and originally derived by \citet{kot15}.
The equation we used to calculate viscosity $\eta$ from the spring damping coefficients $\gamma_s$
 is equation 31 by \citet{frouard16}.
Our simulated material is compressible, so energy damping arises from both deviatoric and volumetric
  stresses.  Springs are created at the beginning of the simulation and do not 
grow or fail during the simulation, so there is no plastic deformation.  

We work with mass in units of $M$,  the mass of the asteroid,
distances in units of volumetric radius, $R_{\rm vol}$, the radius of a spherical body with the same
volume, and time in units of $t_{\rm grav}$  (equation \ref{eqn:tgrav}). 
\begin{align}
t_{\rm grav} &\equiv \sqrt{\frac{R_{\rm vol}^3}{GM}} = \sqrt{ \frac{3}{4 \pi G \rho}} \nonumber \\
&= 1685 {\rm s} \left(\frac{\rho}{1260\ {\rm kg\ m}^3} \right)^{-\frac{1}{2}}. \label{eqn:tgrav}
\end{align}
The density in these units is 
\begin{equation} 
\rho = \frac{3}{4 \pi}. \label{eqn:rho}
\end{equation}
A convenient unit of energy density
\begin{equation}
e_{\rm grav} = \frac{GM^2}{R_{\rm vol}^4} = 110 \ {\rm kPa} \left( \frac{M}{7.8 \times 10^{10} {\rm kg}} \right)^2
\left(\frac{R_{\rm vol}}{246 \ {\rm m}} \right)^{-4}. \label{eqn:eg}
\end{equation}
Elastic moduli, such as shear modulus $\mu$ and Young's modulus $E$, are given in units of $e_{\rm grav}$ (equation \ref{eqn:eg})
which scales with the gravitational energy density or central pressure.
All mass nodes have the same mass
and all springs have the same spring constants.
For numerical stability,
the simulation time-step must be chosen so that it is shorter than the time it takes elastic waves to travel
between nodes \citep{frouard16,quillen16_haumea}.

Initial node distribution and spring network is chosen with 
 the triaxial ellipsoid random spring model
described by \citet{frouard16,quillen16_haumea}.
The surface obeys $\frac{x^2}{a^2} + \frac{y^2}{b^2} + \frac{z^2}{c^2} = 1$
with $a,b,c$ equal to half the lengths of the principal body axes.  An oblate ellipsoid is described
with an axis ratio $c/a<1$ and  $b=a$ and has a pancake-like shape.   A prolate ellipsoid has
$c/a <1$ and $c=b$ and has a cigar-like shape.  Oblates and prolates are sometimes described as `biaxial' or as `ellipsoids of revolution'.
Particle positions within a cube containing the body's surface are drawn from a uniform spatial distribution
but only accepted as mass nodes into the spring network if they are within the shape model
and if they are more distant than $d_I$ from every other previously generated particle.
Once the particle positions have been generated, every pair of particles within distance $d_s$ of each
other are connected with a single spring.  The parameter $d_s$ is the maximum rest
length of any spring.   
We chose $d_s/d_I$ so that the number of springs per node is greater than 12, as recommended
by \citet{kot15}, so as to better approximate a homogenous and isotropic elastic solid.  Springs are initiated at their rest lengths.
The body is not initially in equilibrium because the body will slightly compress due to self-gravity.
(All mass nodes exert gravitational forces on each other during the simulation.)
We begin each simulation with an increased damping parameter.  The simulations
are run for a time $t_{\rm damp}$ to remove vibrations as the body settles into an equilibrium state.
After this damping period  we set the spring damping parameter to the desired value (for the simulation).
Measurements are only done on the simulation output after this time.

\begin{figure}
\centering
$\begin{array}{c}
\includegraphics[width=3.0in, trim={0mm 0mm 0mm 0mm},clip]{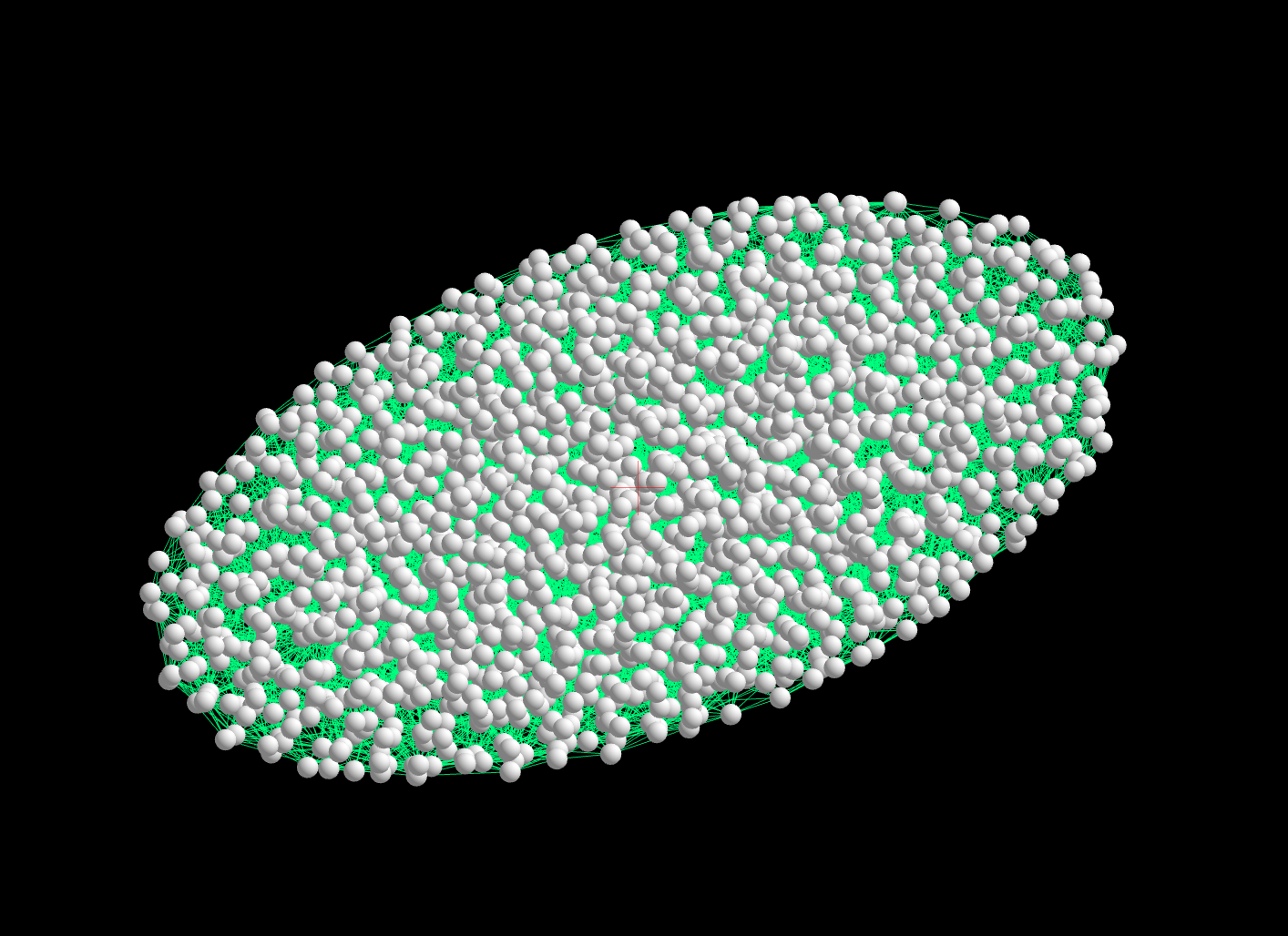} \\
\includegraphics[width=3.0in, trim={0mm 0mm 0mm 0mm},clip]{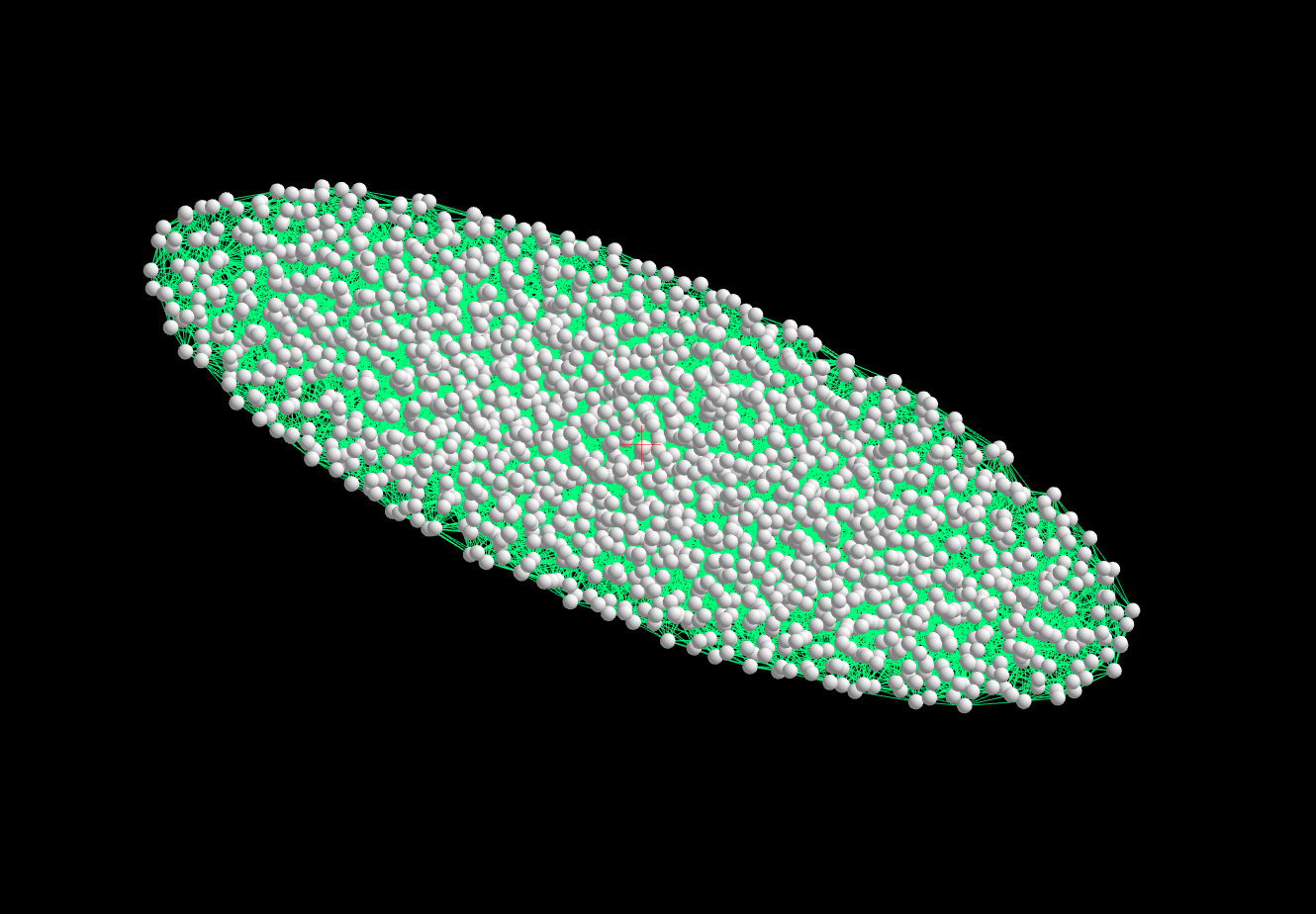}  
\end{array}$
\caption{Top panel:  A simulation snap-shot of an oblate spinning body.
Bottom panel: A simulation snapshot of a prolate spinning body.
The oblate axis ratio is $a:c =$ 1:0.33 with $a=b$ and the prolate axis ratio is 
$a:c = $ 1:0.33 with $b=c$. These are snapshots taken from the simulations while they were running.
The angular momentum vector points upward and they are undergoing non-principal axis rotation (NPA).
\label{fig:snaps}}
\end{figure}

\begin{table}
\vbox to 125mm
{
        \centering
        \caption{\large Some Nomenclature}
        \label{tab:nomen}
\begin{tabular}{lllll} 
\hline
Non-principal axis   & NPA \\
Damping timescale  & $\tau$ \\
Asteroid mass           &      $M$                \\
Radius of equivalent volume & $R_{\rm vol}$ \\
Mass density & $\rho $ \\
Semi-axes of ellipsoid  & $a,b,c$ \\
Body axis ratio if oblate or prolate      & $h$                \\
Spin vector & $\boldsymbol \Omega$ \\
Initial spin vector & ${\boldsymbol \Omega}_0$ \\
Wobble or NPA angle & $\theta$ \\
Angle between ${\boldsymbol \Omega}$ and axis of symmetry & $ \alpha $ \\
Moment of inertia parallel to body symmetry axis & $I_{\parallel}$ \\
Moment of inertia perpendicular to body symmetry axis & $I_{\perp}$ \\
Angular momentum vector & $ \bf J$ \\
Final spin if oblate & $\tilde \omega $ \\
Precession rate in body frame & $\omega_{\rm prec}$ \\
Shear modulus & $ \mu$ \\
Young's modulus & $E$ \\
Viscosity & $\eta$ \\
Viscoelastic relaxation time & $\tau_{\rm relax}$ \\
Poisson ratio & $\nu$ \\
Kelvin-Voigt viscoelastic model & KV \\
Maxwell viscoelastic model & MW \\
\citet{frouard18} & FE \\
\citet{breiter12} & BR \\ 
Quality factor & $Q$ \\
Stress  & $\sigma$ \\
Strain & $\epsilon$ \\
\hline
\end{tabular}\\
Notes: $\tilde \omega$ is not equal to the final spin value (when principal axis rotation is reached)
if the body is prolate. \\
}
\end{table}

\subsection{Body Frame Precession frequency for homogeneous oblate and prolate ellipsoids}
\label{sec:freq}

Acceleration in the body frame depends on the precession frequency.   We outline notation
for computation of this frequency for ellipsoids of rotation following \citet{sharma05,frouard18}.
In the body frame, our coordinate system is defined by the three principal axes of inertia 
with coordinates $x, y, z;$ and with unit vectors ${\bf e}_x, {\bf e}_y, {\bf e}_z$.
The angular velocity is denoted by ${\boldsymbol \Omega}$ and the spin angular momentum is $\bf J$.
The motion of a free rigid body  is given by Euler equations
\begin{equation}
I_i \dot \Omega_i = (I_j - I_k) \Omega_j \Omega_k
\end{equation}
with $(i,j,k) = (x,y,z)$ and cyclic transpositions  $(i,j,k) = (y,z,x)$ and  $(i,j,k) = (z,x,y)$. 

We consider 
a free spinning homogeneous oblate or prolate ellipsoid of revolution with semi-axes $a,b,c$.  
For an oblate $a=b>c$ and for a prolate $a>b=c$. 
We assume that  the body's axis of symmetry  is aligned with ${\bf e}_z$, with moment
of inertia $I_\parallel$.  
For prolate and oblate ellipsoids, 
\begin{align}
I_\parallel &= \frac{2 M a^2}{5} \qquad {\rm oblate} \nonumber \\
I_\parallel &= \frac{2 M c^2}{5} \qquad {\rm prolate} \label{eqn:Ipar}  
\end{align}
The other two axes have the same moment of inertia $I_\perp$. 
We use axis ratio $h$ with convention
\begin{align}
h &= \frac{c}{a}  < 1 \qquad {\rm oblate}  \nonumber \\
h &= \frac{a}{c} > 1\qquad {\rm prolate} ,
\end{align}
giving
\begin{align}
I_\perp &=  \frac{1 + h^2}{2} I_\parallel  \label{eqn:Iperp} 
\end{align}
for both prolate and oblate ellipsoids.

In the body frame,
the  angular momentum vector $\bf J$ and spin vector $\boldsymbol \Omega$ are 
both precessing with frequency $\omega_{\rm prec}$ about the body's axis of symmetry. 
Euler's equations give a constant $\Omega_z$  and
\begin{align}
\dot {\boldsymbol \Omega} &= -\left(\frac{1-h^2}{1+h^2} \right) \Omega_z
 \left [ -\Omega_y  {\bf e}_x  + \Omega_x  {\bf e}_y \right] . \label{eqn:Omdot}
\end{align}
The factor $-(1-h^2)/(1+h^2) = (I_\perp - I_\parallel)/I_\perp$ and    
\begin{align}
\Omega_x &= \Omega \sin \alpha \cos (\omega_{\rm prec} t ) \nonumber \\
\Omega_y &= \Omega \sin \alpha \sin (\omega_{\rm prec} t ) \nonumber  \\ 
\Omega_z &= \Omega \cos \alpha = \frac{J}{I_\parallel} \cos \theta   \label{eqn:Omegaz}
\end{align}
where $\alpha$ is the angle between spin $\boldsymbol \Omega$ and body axis ${\bf e}_z$,
and $\theta$ is the angle between angular momentum $\bf J$ and body axis ${\bf e}_z$. 
This angle is sometimes called the non-principal angle or wobble angle
 and it is one of the Andoyer-Deprit canonical variables \citep{celletti10}.
The angle  $\alpha$ and NPA angle $\theta$  are related by
\begin{equation}
\tan \theta = \frac{I_1}{I_3} \tan \alpha = \frac{1 +  h^2}{2} \tan \alpha. \label{eqn:tanalpha}
\end{equation}

We define the frequency (following \citealt{sharma05,frouard18})
\begin{align}
\tilde \omega &\equiv \frac{J}{I_\parallel} \label{eqn:tilde_omega}. 
\end{align}
Our $\tilde \omega$ is equivalent to  $\Omega_0$ defined by \citet{sharma05} and $\tilde \omega_3$ defined by \citet{frouard18}. 
This is the final spin rate of a homogenous oblate ellipsoid that has damped
down to principal axis rotation while conserving angular momentum.
Equation \ref{eqn:Omdot} implies that the precession frequency (for both prolate and oblate ellipsoids)
\begin{align}
\omega_{\rm prec} &= \left( \frac{I_\parallel} {I_\perp} - 1 \right) \Omega_z = 
\left( \frac{1 - h^2}{1+h^2}  \right) \frac{J}{I_\parallel} \cos \theta  \nonumber \\
&= \left( \frac{1 - h^2}{1+h^2}  \right)  \cos \theta\  \tilde \omega.
\label{eqn:omega_prec}
\end{align}
The frequency $\omega_{\rm prec}$ is the precession
rate of spin vector $\boldsymbol \Omega$ and angular momentum vector $\bf J$ in the body frame.
The frequency $\Omega_z$ (equation \ref{eqn:Omegaz}) 
is equivalent to  $ \Omega_3$ used by \citet{frouard18}. 

Using equations \ref{eqn:Omegaz}, \ref{eqn:tanalpha}, \ref{eqn:tilde_omega} and \ref{eqn:omega_prec}
it is convenient to write spin components in terms of $\tilde \omega$ and NPA angle $\theta$
\begin{align}
\Omega_x&=  \tilde \omega \frac{2}{1 + h^2} \sin \theta\  \cos \omega_{\rm prec} t    \label{eqn:omx_t} \\
\Omega_y  &=  \tilde \omega \frac{2}{1 + h^2} \sin \theta\  \sin \omega_{\rm prec} t \\
\Omega_z &= \tilde \omega \cos \theta.  \label{eqn:omz_t} 
\end{align}
We will use these expressions to generate initial conditions for our simulations.


Here we have defined $\theta$ in terms of angle between angular
momentum and the body's axis of symmetry
not the principal axis with the largest moment of inertia.  For oblate bodies
there is no difference between the two definitions.  However  prolate bodies
reach principal axis rotation at $\theta = \pi/2$ rather than $\theta=0$.
The precession frequency in the body frame ($\omega_{\rm prec}$) should not be confused with precession
frequencies of Euler angles.  

\subsection{Measurements of Energy Dissipation rates}

We run series of short simulations  to measure energy dissipation rates for wobbling spinning bodies.
To measure the dissipation rate in each numerical simulation we compute at each time-step the total energy.
This is a sum of gravitational potential energy, the kinetic energy  and the spring potential energies.
We checked that without dissipation,  the 
sum of gravitational potential energy, elastic potential energy
and kinetic energy (including vibrational and rotational motions) was conserved.
The simulations were integrated for a total time $t_{\rm max}$ which we chose to be between 100 and 400
(in our numerical units of $t_{\rm grav}$).   
For each simulation we plotted total energy versus time.   In most cases the points lie on a line with energy decreasing
with time and with slope giving the measurement of energy dissipation rate in gravitational units.
However,
bodies with slow precession rates took longer to reach an equilibrium state where the energy as a function of time
was linearly decreasing.    We discarded earlier outputs, retaining only
energy measurements after the body has reached an equilibrium state and the energy as a function of time
was linear.  In many cases we saw low amplitude oscillations (in energy) at twice the body precession frequency along with
a steady decrease in energy.  For these we  increased  the total
integration time  so that it was more than a few times the precession frequency, making it possible
to  measure the energy decay rate slope.   In all cases we measured slopes with a linear fit to the energy as a function of time. 
Each simulation took 10 to 25 minutes on a 2018 MacBook Pro with a 2.6 GHz Intel Core i7 processor.

Figure \ref{fig:snaps} shows snap shots of two of our simulations.
The top snapshot shows an oblate ellipsoid  and the bottom one a prolate ellipsoid.
The angular momentum axis in both figures is upward.
The green lines connecting the particles represent the springs.
The gray spheres represent the point mass particle nodes.
These illustrate the number of particles and springs in the simulations discussed in this manuscript.

Table \ref{tab:common} lists common simulation parameters.
Parameters for individual simulations are listed in the subsequent tables. 


We checked that precession rate in the body frame measured from the simulation
 was equal to that given by formula \ref{eqn:omega_prec} for oblate and prolate
bodies.   We keep $\omega_{\rm prec} \tau_{\rm relax} <1$ to remain
in the low frequency viscoelastic regime.
If we go in the other direction $\omega_{\rm prec} \tau_{\rm relax} >1$ then the Kelvin Voigt model would resemble
the Maxwell viscoelastic model.
We prefer to work in the lower limit 
as this gives a model
with weak rather than strong damping in each spring.

\begin{table*}
\vbox to 80mm
{
        \centering
        \caption{\large Common Simulation Parameters}
        \label{tab:common}
\begin{tabular}{llll} %
\hline
Common simulation parameters   \\
\hline
Radius of equivalent volume  & $R_{\rm vol} $ & 1 \\
Mass & $M$ & 1 \\
Density & $\rho$  & $\frac{3}{4 \pi}$ \\
Gravitational timescale  & $t_{\rm grav}$ & 1 \\
Timestep & $dt$          & 0.005 \\
Minimum distance between mass nodes & $d_I$ & 0.12 \\
Ratio of max spring length to $d_I$  & $d_s/d_I$  &  2.3 \\
Number of nodes &  $N_{\rm nodes}$ & $\approx 1800$ \\
Number of springs per node & $N_{\rm springs}/N_{\rm nodes}$ & $\approx 13$ \\
Damping time & $t_{\rm damp}$ & 10 \\
Simulation time & $t_{\rm max}$ & 100--400 \\
\hline
\end{tabular}\\
Note: Units in this table and subsequent tables are in N-body or gravitational units, as described in section \ref{sec:sim_description}.\\
}
\end{table*}

\section{Simulations of Oblate and Prolate ellipsoids}
\label{sec:ob_pro}

We describe a series of simulations that varies body axis ratios and spin axes with the goal
of checking  equation 57  \citet{frouard18}  and the similar equation 103  by \citet{breiter12})
 for the power dissipated by
a homogenous viscoelastic oblate body undergoing NPA rotation.  
 In section \ref{sec:tau} we discuss
how the energy dissipation rate depends on the viscoelastic relaxation timescale.
In section \ref{sec:omega} we explore how the energy dissipation rate depends depends on spin rate.
In section \ref{sec:oblate} and \ref{sec:prolate} we explore how the dissipation rates 
are sensitive to NPA angle $\theta$ and axis ratio $h$ for  oblate and prolate ellipsoids.

\begin{table}
\vbox to 75mm
{
 \caption{\large Oblate ellipsoid simulation parameters for varying viscoelastic relaxation time}
        \label{tab:tau}
\begin{tabular}{llll} %
\hline
Body shape && Oblate \\
Axis ratio & $h$ & 1/3 \\
Spring constant & $k_s$ & 0.08 \\
Shear modulus & $\mu$  &  $ 1.5$ \\ 
Spring damping coefficients & $\gamma_s$ & 1/4, 1/2, 1, 2, 4, 8, 16 \\ 
NPA angle & $\theta$   & $39.3^\circ$ \\
Initial spin  & ${\boldsymbol \Omega}_0$ & (0.3, 0.0, 0.3) \\
Precession rate & $\omega_{\rm prec}$ & 0.243 \\
Frequency   & $\tilde \omega $ &  0.343 \\
\hline
Relevant figure & & Figure \ref{fig:tau} \\
\hline
\end{tabular}\\
Notes: We carried out a series of simulations with different spring damping coefficients.
These simulations are discussed in section \ref{sec:tau}. 
Additional simulation parameters are listed in Table \ref{tab:common}. 
}
\end{table}

\begin{figure}
\centering
\includegraphics[width=3.25in, trim={0mm 0mm 0mm 0mm},clip]{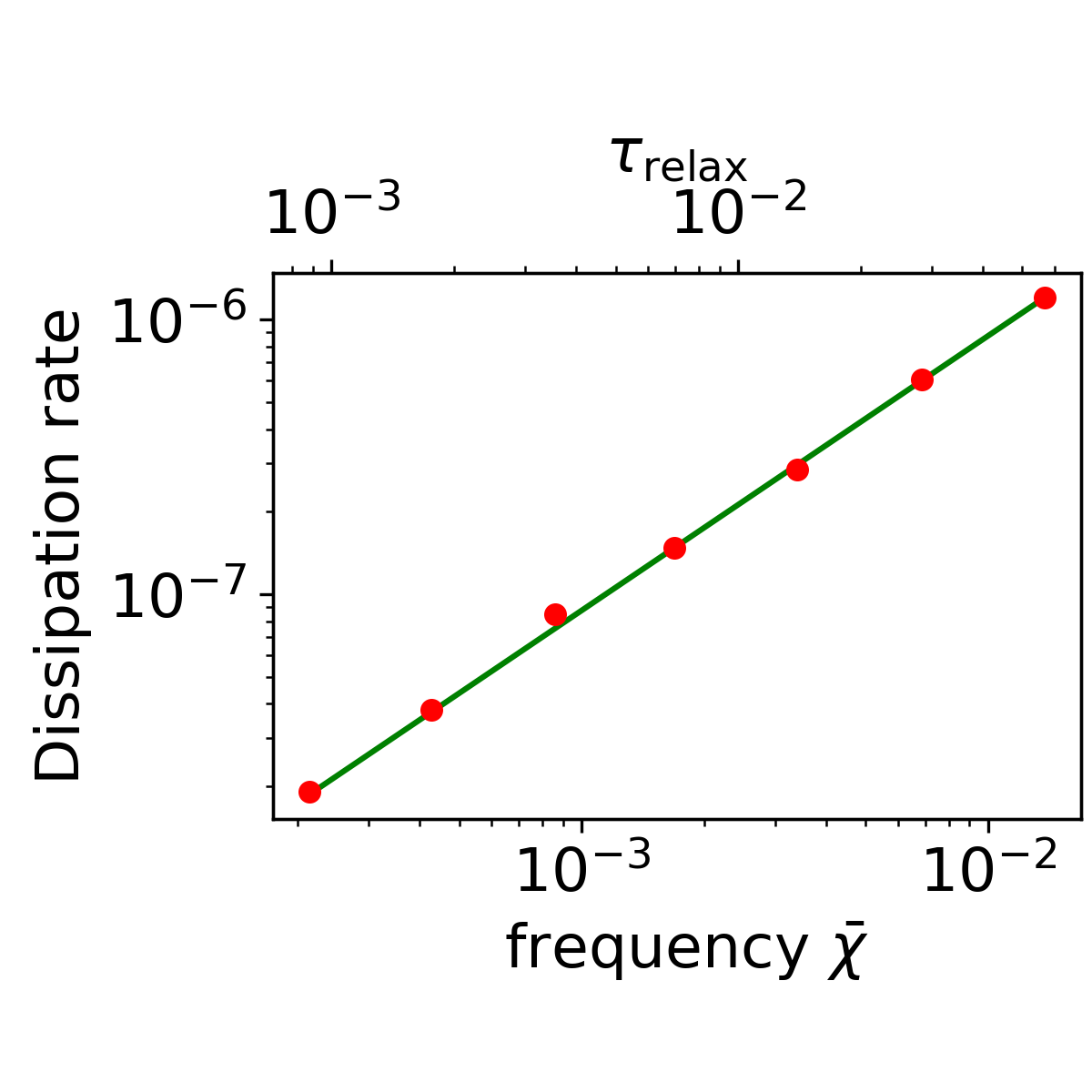}   
\caption{Dependence of dissipation rate during NPA rotation depends on viscoelastic relaxation time.
We show energy dissipation rate measured for a series of simulations of wobbling oblate ellipsoids
with parameters listed in Tables \ref{tab:common} and \ref{tab:tau}.
The simulations are identical except the spring dissipation parameter $\gamma_s$ is varied.
Numerical measurements are shown as red points.  The bottom $x$ axis shows the frequency $\bar\chi = \tau_{\rm relax} \omega_{\rm prec}$,
the top $x$ axis shows viscoelastic relaxation time $\tau_{\rm relax}$ and the $y$ axis shows
 the dissipated power.  The green line is linear in $\tau_{\rm relax}$ or $\bar \chi$, as predicted for
 the Kelvin-Voigt viscoelastic model (with a slope of 1 on the log-log plot) 
 and goes through the point
on the lower left.  The line was not fit to the measured dissipation rates but shows that the points are 
consistent with energy dissipation rate  linear dependent on viscoelastic relaxation time $\tau_{\rm relax}$ or
frequency $\bar \chi$, predicted for the Kelvin-Voigt model for small $\bar \chi$.
}
\label{fig:tau}
\end{figure} 

\subsection{Sensitivity of the energy dissipation rate to viscoelastic relaxation timescale in the Kelvin-Voigt model}
\label{sec:tau}
 
In continuum mechanics,  an elastic/viscoelastic analogy, referred to as the {\it correspondence principle},  
 establishes a relation between frequency dependent stress and strain relations and the 
equivalent static relation 
by replacing the elastic shear modulus with a complex shear modulus (or, equivalently, by replacing the real shear compliance with a complex compliance) and by simultaneous replacing the real strain and stress with their Fourier harmonics.
\citet{frouard18} computed the energy dissipation rate for wobbling
oblate ellipsoids with a Maxwell rheology.  The Maxwell rheology has complex shear 
rigidity (inverse of the complex compliance)
as a function of frequency  $\chi$
\begin{equation}
\tilde \mu(\chi)^{\rm MW} = \mu \frac{i \chi \tau_{\rm relax}}{1 + i \chi \tau_{\rm relax}} = \mu \frac{i \bar \chi}{1 + \bar \chi},  \label{eqn:muMW}
\end{equation}
where 
$\tau_{\rm relax} = \eta/\mu$ is the viscoelastic relaxation timescale and $\bar \chi \equiv \chi \tau_{\rm relax}$
(see appendix D by \citealt{frouard18} for an introduction to the {\it correspondence principle}).
With stress $\sigma = \sigma_0 \cos ( \chi t)$, the  energy dissipation rate per unit volume averaged over the oscillation period
\begin{equation}
 p^{\rm MW}  = \frac{1}{4 \eta} \sigma_0^2, \label{eqn:pMW}
 \end{equation}
 (equation  D7 by \citealt{frouard18})
and is independent of frequency. 
As a result, their estimate for the total power dissipated for a wobbling oblate viscoelastic ellipsoid
(equation 57 by \citealt{frouard18})
  \begin{align}
 P^{\rm FE,MW}_{\rm oblate} = \frac{a^7 {\tilde \omega}^4 \rho^2}{\eta} 
 \left( f^{\rm FE}_{(1)}(h) \sin^2 \theta  \cos^2 \theta+   f^{\rm FE}_{(2)}(h) \sin^4 \theta \right) , \label{eqn:P57}
 \end{align}    
 and does not directly depend on the precession frequency.
Their functions
\begin{align}
f^{\rm FE}_{(1)}(h) &= 
\frac{\pi}{(h^2+1)^4} \frac{32}{315} (h^2 + 1)^2 h^5 \ \times  \nonumber \\
& \  \ \Big[ \frac{(1050 h^4 + 2015h^2 + 507)}{(20h^2 + 13)^2}  \Big] \nonumber \\   
f^{\rm FE}_{(2)}(h) &= 
\frac{\pi}{(h^2+1)^4} \frac{16}{315} h \ \times \nonumber \\
&
\ \  \Big[ \frac{384 h^8 + 960h^6 + 1900h^4 + 1650h^2  + 1125}{(8h^4 + 10h^2 + 15)^2} \Big] .  
\label{eqn:f_FE}
\end{align}   
 
In contrast to equation \ref{eqn:muMW}, the Kelvin-Voigt viscoelastic model has complex shear modulus  
\begin{equation}
\tilde \mu(\chi)^{\rm KV} = \mu (1 + i \bar \chi) \label{eqn:muKV}
\end{equation}
and time averaged energy dissipation rate per unit volume
\begin{equation}
 p^{\rm KV}   = \frac{1}{4 \eta} \sigma_0^2 \frac{{\bar \chi}^2}{1 + {\bar\chi}^2} \label{eqn:pKV}
\end{equation}
(these are equations D3 and D8 by \citealt{frouard18}).
The viscosity $\eta = \mu \tau_{\rm relax}$ for both Maxwell and Kelvin-Voigt viscoelastic models.
 In the limit of small ${\bar \chi} <1 $
\begin{equation}
 p^{\rm KV} \approx p^{\rm MW} \bar \chi^2.  \label{eqn:barchi2}
\end{equation}

The acceleration in the inertial frame is related to that in the body frame by
\begin{equation}
\frac{d{\bf a}}{dt}_{\rm inertial} =  \frac{d{\bf a}}{dt}_{\rm body} + \dot{\boldsymbol\Omega} \times {\bf r} 
	+ {\boldsymbol \Omega} \times ({\boldsymbol \Omega} \times {\bf r})  \label{eqn:acc}
\end{equation}
at position $\bf r$.
As explained by \citet{frouard18}, for an oblate or prolate body, the time dependence of $\boldsymbol \Omega$ depends on the
precession frequency, $\omega_{\rm prec}$.   
However  equation \ref{eqn:acc} shows that the acceleration contains Fourier components that depend on frequencies
  $\omega_{\rm prec}$ and $ 2 \omega_{\rm prec}$.
Using stress components calculated by \citet{sharma05}, 
\citet{frouard18} integrated the time average of stress and stress for both Fourier frequencies
and added the result to compute the total dissipated power.  This accounts for the
two terms in Equation \ref{eqn:f_FE} (from equation 57 by \citealt{frouard18}).
Thus
\begin{equation}
P^{\rm FE,MW}_{\rm oblate}  = P^{\rm FE}_{ (1),{\rm oblate} }+ P^{\rm FE }_{(2),{\rm oblate}}
\end{equation}
where we have verified (by examining terms in the appendices by \citet{frouard18})
that the first term arises from the Fourier components of acceleration 
that depend on frequency  $\omega_{\rm prec}$
(and is computed with $f^{\rm FE}_{(1)}(h)$) 
and the second term arises from those that depend on frequency  $2\omega_{\rm prec}$
(and is computed with $f^{\rm FE}_{(2)}(h)$) .

Using frequencies ${\bar \chi} = \omega_{\rm prec} \tau_{\rm relax} $  and twice this, 
in the limit of small ${\bar \chi} <1 $,  and using equations \ref{eqn:barchi2} and \ref{eqn:P57},
we estimate  the total dissipated power for the Kelvin-Voigt model
\begin{align}
P^{\rm FE,KV}_{\rm oblate} & \approx \left( P^{\rm FE}_{ (1),{\rm oblate}} +4 P^{\rm FE}_{(2),{\rm oblate}} \right) \bar \chi^2  \nonumber \\
&=  \frac{a^7 {\tilde \omega}^4 \rho^2}{\eta}  \omega_{\rm prec}^2 \tau_{\rm relax}^2 \times \nonumber \\
& \qquad
\left( f^{\rm FE}_{(1)}(h)\sin^2\theta \cos^2 \theta + 4 f^{\rm FE}_{(2)} (h)     \sin^4\theta \right)
 .
\label{eqn:PKV}
\end{align}
Note the factor of 4 on $f^{\rm FE}_{(2)} (h)$ arising from $2 \bar\chi$. 

We compare bodies with the same shear modulus $\mu$, body axis ratios, and initial spin vector
but different viscosities.
The dissipated power (equation \ref{eqn:PKV})
\begin{align}
P^{\rm KV} \propto  \frac{\tau_{\rm relax}}{\mu}, \label{eqn:propgamma}
\end{align}    
where we have used viscosity $\eta = \mu \tau_{\rm relax}$. 
If we  change the relaxation time $\tau_{\rm relax}$ in our simulations without changing 
the shear modulus, $\mu$, the body
axis ratio, $h$, the  angle $\theta$ or the initial body spin, $\boldsymbol \Omega$, and we maintain 
$\omega_{\rm prec} \tau_{\rm relax} <1 $, then we expect
to measure a linear relation between the viscoelastic relaxation timescale $\tau_{\rm relax}$
and the energy dissipation rate (as expected for the Kelvin-Voigt model).   We can vary  $\tau_{\rm relax}$ without varying
these other quantities by adjusting the spring damping coefficients $\gamma_s$ and leaving other simulation
parameters fixed.

Figure \ref{fig:tau} shows a series of simulations of oblate bodies with all parameters identical
except they have different values of spring dissipation parameter $\gamma_s$ and 
so they have different relaxation timescales $\tau_{\rm relax}$ and normalized frequencies $\bar\chi$.
For these simulations
the precession rates  satisfy $\omega_{\rm prec} \tau_{\rm \relax} <1$ so as to remain
in the  regime where $\bar \chi<1$ for equation \ref{eqn:pKV} and giving an expected linear
dependence of power on  $\tau_{\rm relax}$ (as in equation \ref{eqn:propgamma}).
The parameters for these simulations are listed in Tables \ref{tab:common} and Table \ref{tab:tau}.
We have plotted energy dissipation rate 
versus frequency $\bar\chi = \omega_{\rm prec} \tau_{\rm \relax}$ (bottom $x$ axis) and viscoelastic
relaxation time $\tau_{\rm relax}$  (top $x$ axis)
in Figure \ref{fig:tau} as red points, measured from for 7 simulations with $\gamma_s$ ranging from 1/4 to 16.  
We show a line in Figure \ref{fig:tau} that has power $P \propto \bar\chi$.  The line was
not fit to the measured points but was adjusted to intersect the point on the lower left.
The green line shows that dissipated power is proportional to $\bar\chi  $,  
$\gamma_s$ and  
 the viscoelastic relaxation time, as expected from equation \ref{eqn:PKV}  (with all other parameters
 remaining fixed).
We confirm that the dissipation rate is proportional to the viscoelastic relaxation time
with the Kelvin-Voigt model. 

\begin{figure}
\centering
\includegraphics[width=3.25in, trim={0mm 0mm 0mm 0mm},clip]{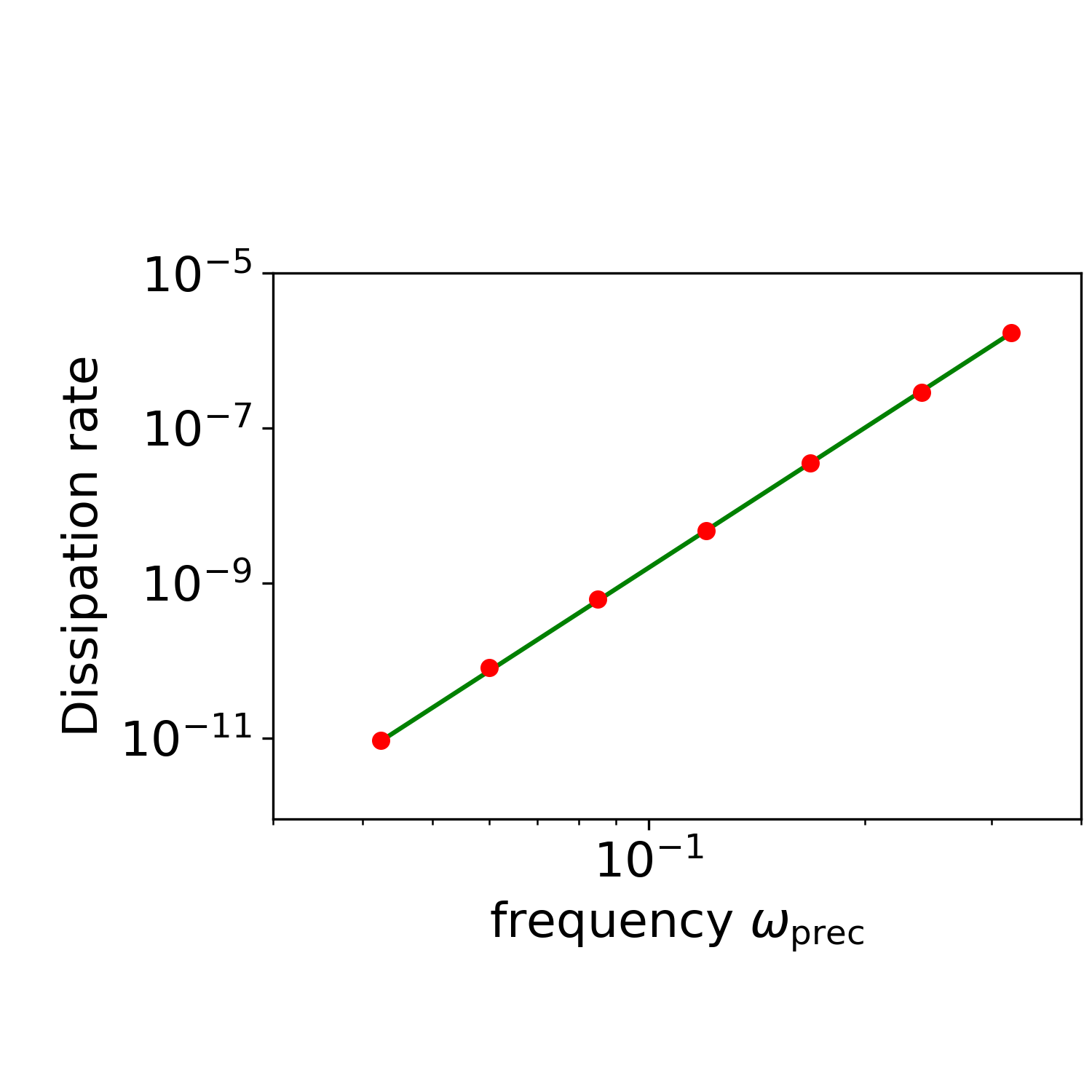}  
\caption{We compare the energy dissipation rates for simulations that are identical 
except for the amplitude of their initial spin vectors $|{\boldsymbol \Omega}_0|$. 
The parameters for these simulations are listed in Tables \ref{tab:common} and Table \ref{tab:omega}.
The $x$ axis is the precession rate $\omega_{\rm prec}$ and the $y$ axis the measured energy dissipation
rate.   The red points show simulation measurements.  The green line has power $P \propto \omega_{\rm prec}^6$, 
predicted for the Kelvin-Voigt model.  It was
 not fit to the points, but was adjusted to intersect the point on the upper right. 
The dissipation rate is sensitive to spin amplitude and with power that depends on
rheology.
}
\label{fig:omega}
\end{figure}

\begin{table}
\vbox to 75mm
{
 \caption{\large Oblate ellipsoid simulation parameters for varying spin amplitude}
        \label{tab:omega}
\begin{tabular}{llll} %
\hline
Body shape && Oblate \\
Axis ratio & $h$ & 1/3 \\
Spring constant & $k_s$ & 0.08 \\
Shear modulus & $\mu$  &  $ 1.5$ \\ 
Spring damping coefficient & $\gamma_s$ & 4 \\ 
Viscoelastic relaxation time & $t_{\rm relax}$ &   0.014 \\
NPA angle & $\theta$   & $39.3^\circ$ \\
Initial spin $x$ components &  ${\Omega}_{0x} $ & $=  {\Omega}_{0z}$    \\
Initial spin  $y$ component & ${\Omega}_{0y}$ & 0   \\
Initial spin  $z$ components & $\Omega_{z0}$ & 0.4, 0.3, 0.21, 0.15, \\
"&& and 0.106, 0.075, 0.053 \\
\hline
Relevant figure && Figure \ref{fig:omega} \\
\hline
\end{tabular}\\
Notes: We carried out a series of simulations with different initial spin values.
These simulations are discussed in section \ref{sec:omega}. 
Additional simulation parameters are listed in Table \ref{tab:common}.
\\
}
\end{table}

\subsection{Sensitivity of the energy dissipation rate to spin rate for oblate ellipsoids}
\label{sec:omega}

Equation \ref{eqn:PKV} gives an
expression for the energy dissipation rate for oblate ellipsoids based on that estimated by  \citet{frouard18}
but modified for a Kelvin-Voigt viscoelastic solid.
Using the expression for the precession frequency in equation \ref{eqn:omega_prec},
 we can write
equation \ref{eqn:PKV} as
\begin{align}
P^{\rm FE,KV}_{\rm oblate} & \approx  \frac{a^7 {\tilde \omega}^6 \rho^2 \tau_{\rm relax}}{\mu}  \frac{(1 -h^2)^2}{(1 + h^2)^2} \cos^2 \theta \ \times 
\nonumber \\ 
& \qquad \left(   f^{\rm FE}_{(1)}(h) \sin^2 \theta \cos^2 \theta+ 4f^{\rm FE}_{(2)}(h) \sin^4 \theta \right)  .
   \label{eqn:PKV2}
\end{align}
We estimate that the dissipated power is strongly dependent on the spin rate or $P \propto \tilde \omega^6$.
We can test this  by comparing simulations with the same NPA angle $\theta$, shear modulus $\mu$, 
viscoelastic relaxation time $\tau_{\rm relax}$, and body axis ratio $h$,
but different initial $|{\boldsymbol \Omega}_0|$ spin amplitudes, 
as $\tilde \omega \propto |{\boldsymbol \Omega}| \propto \omega_{\rm prec}$ at fixed $\theta$; see equation 
\ref{eqn:omega_prec}.
Figure \ref{fig:omega} shows such a series of oblate ellipsoid simulations.
 The simulation parameters are listed in Tables \ref{tab:common}
and \ref{tab:omega}. We plot as red dots energy dissipation rate measured from the simulations 
but against the spin precession rate $\omega_{\rm prec}$ which is proportional to $\tilde \omega$.
We again show a line on the plot that is not fit to the simulation
measurements but depends on  $\omega_{\rm prec}^6$.  
The line on the figure shows that the energy dissipation rate is consistent
with being proportional to $\tilde \omega^6$ as expected
from equation \ref{eqn:PKV2}.

We find that the Kelvin-Voigt model has power proportional to spin $\tilde \omega$ to the sixth power, 
whereas the Maxwell model has power proportional to the fourth power (equation \ref{eqn:P57} and that predicted by \citealt{frouard18}).   
With a quality factor model, the energy dissipation rate is proportional to the fifth power (equation 81 by  \citealt{breiter12}).
Wobble damping times are  sensitive to rheology.

\begin{table}
\vbox to 80mm
{
\caption{\large Oblate  ellipsoid simulation parameters for varying axis ratio and NPA angle}
        \label{tab:oblate}
\begin{tabular}{llll} %
\hline
Body shape && Oblate \\
Spring constant & $k_s$ & 0.08 \\
Shear modulus & $\mu$  &  $ 1.5$ \\ 
Spring damping coefficients & $\gamma_s$ & 4 \\ 
Viscoelastic relaxation time & $t_{\rm relax}$ &   0.014 \\
Frequency & $\tilde \omega$ & 0.5 \\
\hline
Axis ratios & $h$ & 1/3, 1/2, 2/3 \\
NPA angles & $\theta$   & 10$^\circ$ to 80$^\circ$ w. increments of 10$^\circ$  \\
\hline
Axis ratios & $h$ & 0.1 to 0.9 w. increments of  0.1   \\
NPA angles & $\theta$   & $30,45,60^\circ$ \\
\hline
Relevant figures && Figures \ref{fig:oblate_FE} and \ref{fig:oblate_BR} \\
\hline
\end{tabular}\\
Notes: We carried out two sets of simulations.  Each simulations has a different value of $h$ and $\theta$.
These simulations are discussed in section \ref{sec:oblate}. 
Initial spin components $\Omega_{x0}, \Omega_{0z}$ were computed using equations \ref{eqn:omx_t} and \ref{eqn:omz_t}
from frequency $\tilde \omega$ and  NPA angle $\theta$.
Additional simulation parameters are listed in Table \ref{tab:common}.
\\
}
\end{table}

\begin{table}
\vbox to 80mm
{
        \caption{\large Prolate ellipsoid simulation parameters for varying axis ratio and NPA angle}
        \label{tab:prolate}
\begin{tabular}{llll} %
\hline
Body shape &&  Prolate\\
Spring constant & $k_s$ & 0.08 \\
Shear modulus & $\mu$  &  $ 1.5$ \\ 
Spring damping coefficients & $\gamma_s$ & 4 \\ 
Viscoelastic relaxation time & $t_{\rm relax}$ &   0.014 \\
Frequency & $\tilde \omega$ & 0.5 \\
\hline
Axis ratios & $h$ &  3/2, 2, 3 \\
NPA angle & $\theta$   & 10$^\circ$ to 80$^\circ$ w. increments of 10$^\circ$\\
\hline
Axis ratios & $h$ & 
1.11, 1.26, 1.50, 1.78, 2.11, 2.51, \\
" & " & and 2.98, 3.54, 4.21, 5.00 \\
NPA angles & $\theta$   & $30,45,60^\circ$ \\
\hline
Relevant figure  & & Figure  \ref{fig:prolate_BR} \\
\hline
\end{tabular}\\
Notes:  We carried out two sets of simulations.  Simulations have different $h$ and $\theta$.
These simulations are discussed in section \ref{sec:prolate}. 
Initial spin components $\Omega_{x0}, \Omega_{0z}$ were computed using equations \ref{eqn:omx_t} and \ref{eqn:omz_t}
from frequency $\tilde \omega$ and  NPA angle $\theta$.
Additional simulation parameters are listed in Table \ref{tab:common}.
\\
}
\end{table}

\begin{figure}
\centering
\includegraphics[width=3.25in, trim={6mm 10mm 0mm 30mm},clip]{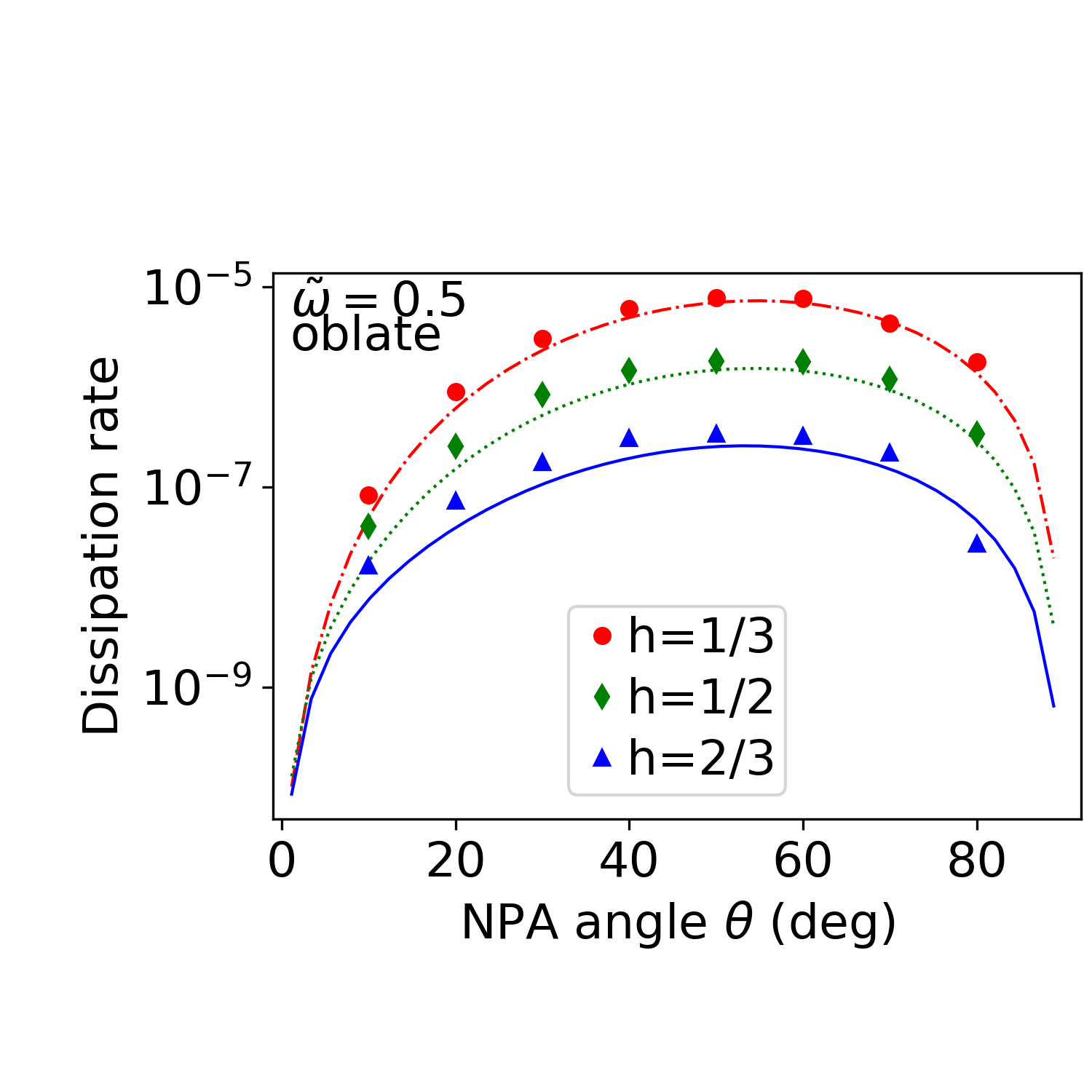}  
\includegraphics[width=3.25in, trim={6mm 10mm 0mm 30mm},clip]{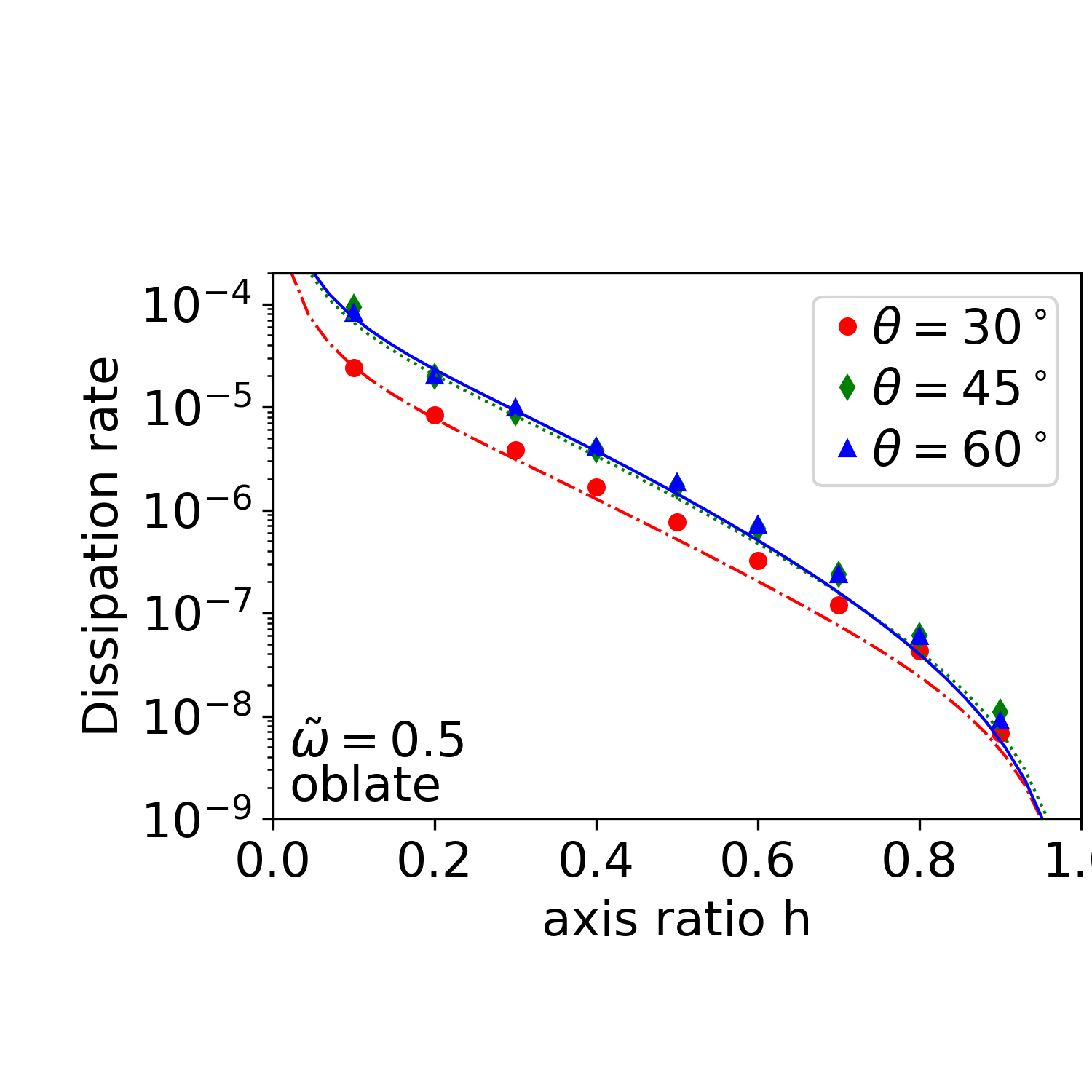}  
\caption{We compare the energy dissipation rates for oblate simulations that are identical 
except for their body axis ratio and NPA angle $\theta$.  
The parameters for these simulations are listed in Tables \ref{tab:common} and Table \ref{tab:oblate}.
The $x$ axis is the NPA angle $\theta$ and the $y$ axis the measured energy dissipation rate.
a) The red circles, green  diamonds and blue triangles show measurements
from simulations of oblate ellipsoids with axis ratios $h=1/3,1/2, 2/3$, respectively.  
The red dash-dotted, green dotted and blue solid lines lines are computed for the same axis ratios 
 using equation \ref{eqn:PKV3_ob} that is based on work by \citet{frouard18}.
b)  Similar to a) except red circles, green  diamonds and blue triangles 
show simulations with  NPA angle  $\theta=30,45, 60^\circ$, and the $x$ axis is the axis ratio $h$.
The red dash-dotted, green dotted and blue solid lines lines are computed for the same NPA angles 
 using equation \ref{eqn:PKV3_ob}.
}
\label{fig:oblate_FE}
\end{figure} 

\begin{figure}
\centering
\includegraphics[width=3.25in, trim={6mm 10mm 0mm 30mm},clip]{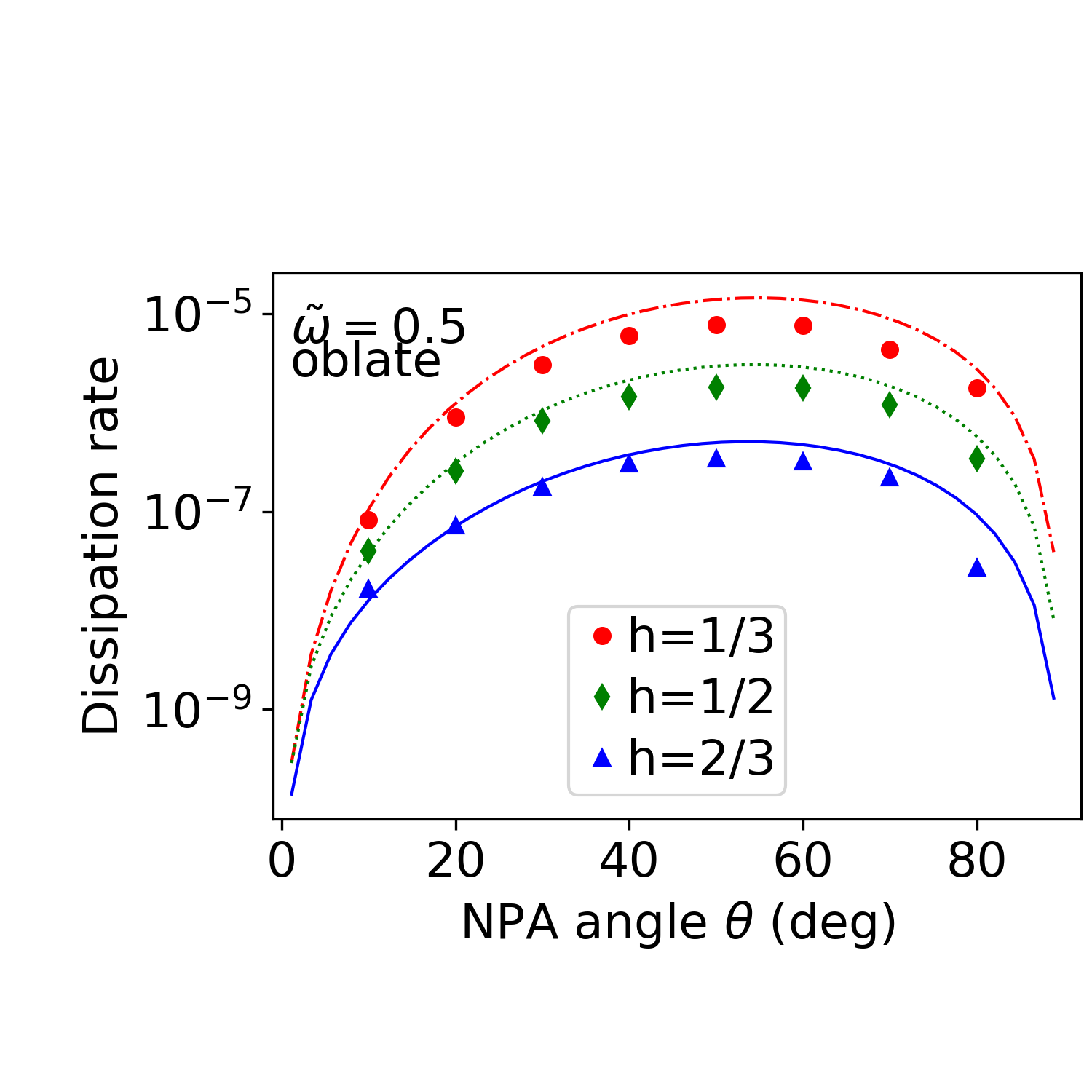}  
\includegraphics[width=3.25in, trim={6mm 10mm 0mm 30mm},clip]{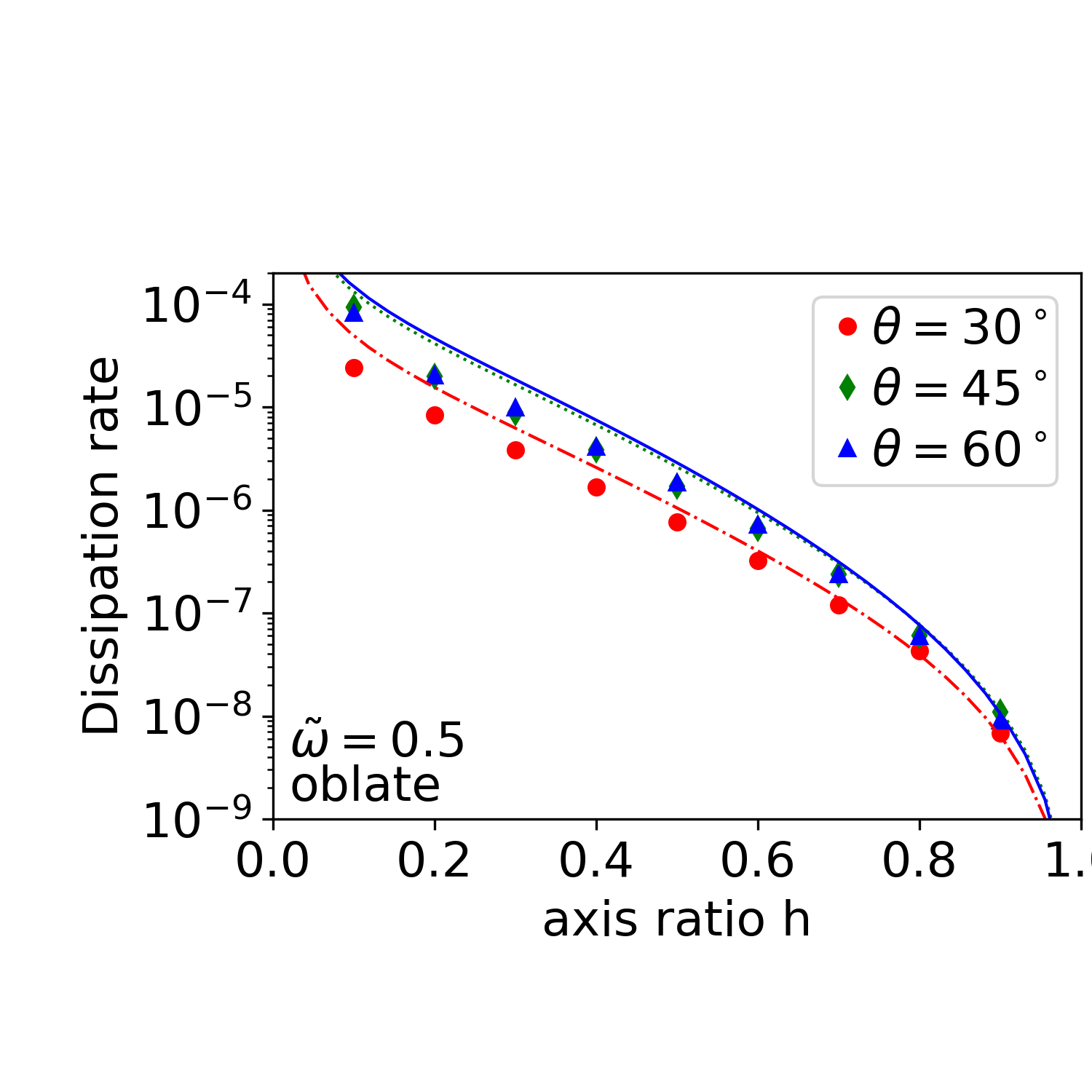}  
\caption{We compare the energy dissipation rates for oblate simulations that are identical 
except for their body axis ratio and NPA angle $\theta$.    These figures are the same as those in Figure \ref{fig:oblate_FE}
except we plot lines computed using equation \ref{eqn:BR3} that are based on calculations
by \citet{breiter12}.
}
\label{fig:oblate_BR}
\end{figure}

\subsection{Sensitivity of dissipation rate to axis ratio and angle for oblates}
\label{sec:oblate}

To explore the sensitivity of the dissipation rate to body axis and spin axis orientation
we fix the frequency $\tilde \omega$, shear modulus and viscosity but vary
the body axis ratio $h$ and  NPA angle $\theta$.
We ran 3 sets of oblate ellipsoid simulations with body axis ratios $h=1/3, 1/2,2/3$.
Each set contains 8 simulations with NPA angles $\theta  = 10,20,30,40,50,60,70$ and $80^\circ$.
We also ran 3 sets of oblate simulations with $\theta = 30, 45, 60^\circ$ and 10 different values of axis ratio $h$.
Initial spin vectors for the simulations were 
set using equations \ref{eqn:omx_t} and \ref{eqn:omz_t} at $t=0$.
Parameters for these simulations are listed in Table \ref{tab:oblate} and the energy dissipations are plotted  
in Figures \ref{fig:oblate_FE} and \ref{fig:oblate_BR} as a function of NPA angle (top panels) and axis ratios $h$ (bottom panels).

The normalization in equation \ref{eqn:P57}  depends on semi-major axis $a^7$.
Our simulations work in units of the radius of the volume equivalent sphere, with  $V = \frac{4}{3} \pi a b c = \frac{4 }{3} \pi R_{\rm vol}^3$.
For an oblate body with semi axes $a=b$, volume equivalent radius $R_{\rm vol}$ and axis ratio $h=c/a$, 
we find that semi-major axis $a= h^{-\frac{1}{3}} R_{\rm vol}$. Inserting this
into equation \ref{eqn:PKV2} we find 
\begin{align}
P^{\rm FE,KV}_{\rm oblate} & \approx    \frac{R_{\rm vol}^7 \rho^2 \tau_{\rm relax} {\tilde \omega}^6}{\mu h^\frac{7}{3} } \frac{(1 -h^2)^2}{(1 + h^2)^2} \cos^2 \theta\    \times  \nonumber \\
& \qquad \left(   f^{\rm FE}_{(1)}(h) \sin^2 \theta \cos^2 \theta+ 4f^{\rm FE}_{(2)}(h) \sin^4 \theta \right)
   \label{eqn:PKV3_ob}
\end{align}
With $\rho, {\tilde \omega}, \mu, \tau_{\rm relax}, R_{\rm vol}$ in our N-body units,
power is in units of $e_{\rm grav} t_{\rm grav}^{-1}$.

In Figure \ref{fig:oblate_FE}  we overplot
lines given by equation \ref{eqn:PKV3_ob} for oblates.   To compute these we used
the shear viscosity and viscoelastic relaxation time estimated from the code (and with parameters
listed in Table \ref{tab:oblate}), $R_{\rm vol}=1$, and a density of $\rho = 3/(4 \pi)$, consistent with our N-body units.
Three lines in the top panel are computed for three axis ratios
$h=1/3, 1/2 $ and 2/3  and shown as red dash-dotted, green dotted and blue solid lines in Figure \ref{fig:oblate_FE}a
as a function of NPA angle $\theta$.
In Figure \ref{fig:oblate_FE}b
three lines are computed for three NPA angles $\theta = 30,45,60^\circ$ and are 
shown as red dash-dotted, green dotted and blue solid lines 
as a function of varying axis ratio $h$.
The model lines are a good match to the numerical measurements, without any additional offsets or normalization.
This implies that the analytical prediction 
by \citet{frouard18} for the Maxwell viscoelastic model is robust, even though we have
modified it for the Kelvin-Voigt model.   

A perfect match between model and numerical measurements is unlikely because
the analytical computation ignores bulk viscosity though our simulated viscoelastic material is compressible
and has a bulk viscosity.   
The bulk viscosity is neglected in the model (equation \ref{eqn:PKV3_ob}) because it is based on calculations by \citet{frouard18} who neglected bulk viscosity. 
The computations by \citet{sharma05},  \citet{frouard18} and \citet{breiter12} assume a Poisson ratio $1/4$ which is 
consistent  with our simulated elastic material.
In the analytical computations displacements and their partial derivatives are assumed to be small.
The analytical models by \citet{sharma05,frouard18} (but not \citealt{breiter12}) ignore compression due to self-gravity.
Stress due to constant inertial acceleration (non-varying components of the acceleration; see
discussion in appendix C by \citealt{frouard18}) was ignored by \citet{frouard18} but not by \citet{sharma05}.

With tidal spin down (but undergoing principal axis rotation) we found that analytical predictions
lacking bulk viscosity were 30\% lower than the numerical measurements \citep{frouard16}.
A factor of 30\% gives $\log_{10} 1.3  = 0.11$  in log space which would look small
in Figure \ref{fig:oblate_FE}.   Many of the numerical measurements  are slightly higher than the predicted values, and this
might be due to the neglect of bulk viscosity in the model that is present in the simulated material.

Using a different derivation of the stress tensor, \citet{breiter12}  derive the energy dissipation
rate for both prolate and oblate ellipsoids.
For ellipsoids of rotation (prolates and oblates), the NPA angle used by \citet{breiter12} 
(their $\theta_s$ defined in their equation 81) 
is equivalent to our 
NPA angle $\theta$. Our precession rate $\omega_{\rm prec}$ is equivalent to their fundamental frequency
of wobbling (defined in their equation 58).  Their frequency $\tilde \omega_s$ (defined in their equation 52) is equivalent to
our $\tilde \omega$.
The energy dissipation rate for oblates  (equations 103,104, 105 by \citealt{breiter12}) can be written as
\begin{align}
P^{\rm BR}_{\rm oblate} &= \frac{a^7 \rho^2  \tilde \omega^5}{\mu Q} \frac{(1 -h^2)}{(1 + h^2)} \cos \theta \ \times \nonumber \\
& \ \ \ \ 
\left( f^{\rm BR}_{(1)} (h) \sin^2 \theta \cos^2 \theta +  f^{\rm BR}_{(2)} (h) \sin^4 \theta \right), \label{eqn:BR1}
\end{align}
with functions
\begin{align}
 f^{\rm BR}_{(1)} (h) &= \frac{32\pi}{105}  \frac{2 h^5}{(1+ h^2)^4} \frac{26 + 35 h^2}{13 + 20 h^2} \\
 f^{\rm BR}_{(2)} (h) &= \frac{32\pi}{105}  \frac{h}{(1+ h^2)^4} \frac{25 + 20 h^2 + 16h^4}{15 + 10 h^2 + 8 h^4} .
\end{align}
They model the energy dissipation with a constant quality factor $Q$.
This effectively gives a frequency dependent viscoelastic relaxation time $\tau_{\rm relax} \approx Q/ \omega_{\rm prec}$ or
\begin{equation}
Q  =  \tau_{\rm relax} \omega_{\rm prec} = \frac{\eta \omega_{\rm prec} }{\mu}. \label{eqn:Q}
\end{equation}  
We can estimate
the energy dissipation rate for a Maxwell model using viscosity $\eta = \tau_{\rm relax} \mu$ and giving
\begin{align}
P^{\rm BR,MW}_{\rm oblate} &= \frac{a^7 \rho^2  \tilde \omega^4}{\eta }  
\left( f^{\rm BR}_{(1)} (h) \sin^2 \theta \cos^2 \theta +  f^{\rm BR}_{(2)} (h) \sin^4 \theta \right), \label{eqn:BR2}
\end{align}
where we have used equation \ref{eqn:omega_prec} for the precession rate and replaced $Q$ 
in equation \ref{eqn:BR1} using equation \ref{eqn:Q}.
This equation closely resembles equation \ref{eqn:P57} (based on equation 57 by \citealt{frouard18}) 
except the $h$ dependent functions
differ.  We identify two terms in equation \ref{eqn:BR2}, 
the first likely arises from acceleration terms that have frequency $\omega_{\rm prec}$ 
and the second  from acceleration terms  dependent on $2\omega_{\rm prec}$. 
We  estimate the energy dissipation rate for the Kelvin Voigt model as we did in section \ref{sec:tau} giving
\begin{align}
P^{\rm BR,KV}_{\rm oblate} &= \frac{R_{\rm vol}^7 \rho^2  \tau_{\rm relax} \tilde \omega^6}{\mu h^{\frac{7}{3}}}  
 \frac{(1 -h^2)^2}{(1 + h^2)^2} \cos^2 \theta\    \times  \nonumber \\
& \quad \left( f^{\rm BR}_{(1)} (h) \sin^2 \theta \cos^2 \theta + 4 f^{\rm BR}_{(2)} (h) \sin^4 \theta \right). \label{eqn:BR3}
\end{align}
Lines computed using equation \ref{eqn:BR3} are plotted on the dissipation rate measurements
for our oblate simulations in Figure \ref{fig:oblate_BR}.   The lines are good match to the numerical measurements.
However, they lie slightly above the numerical measurements rather than slightly below them as did those by \citet{frouard18}.
Within a factor of about 20\%, the quality factor based dissipation model by \citet{breiter12} is consistent our numerical measurements and
with the Maxwell model computed by \citet{frouard18} for oblates if we use equation \ref{eqn:Q} to relate $Q$ to the
viscoelastic relaxation time or viscosity.

\subsection{Sensitivity of dissipation rate to axis ratio and angle for prolates}
\label{sec:prolate}

\begin{figure}
\centering
\includegraphics[width=3.25in, trim={6mm 10mm 0mm 30mm},clip]{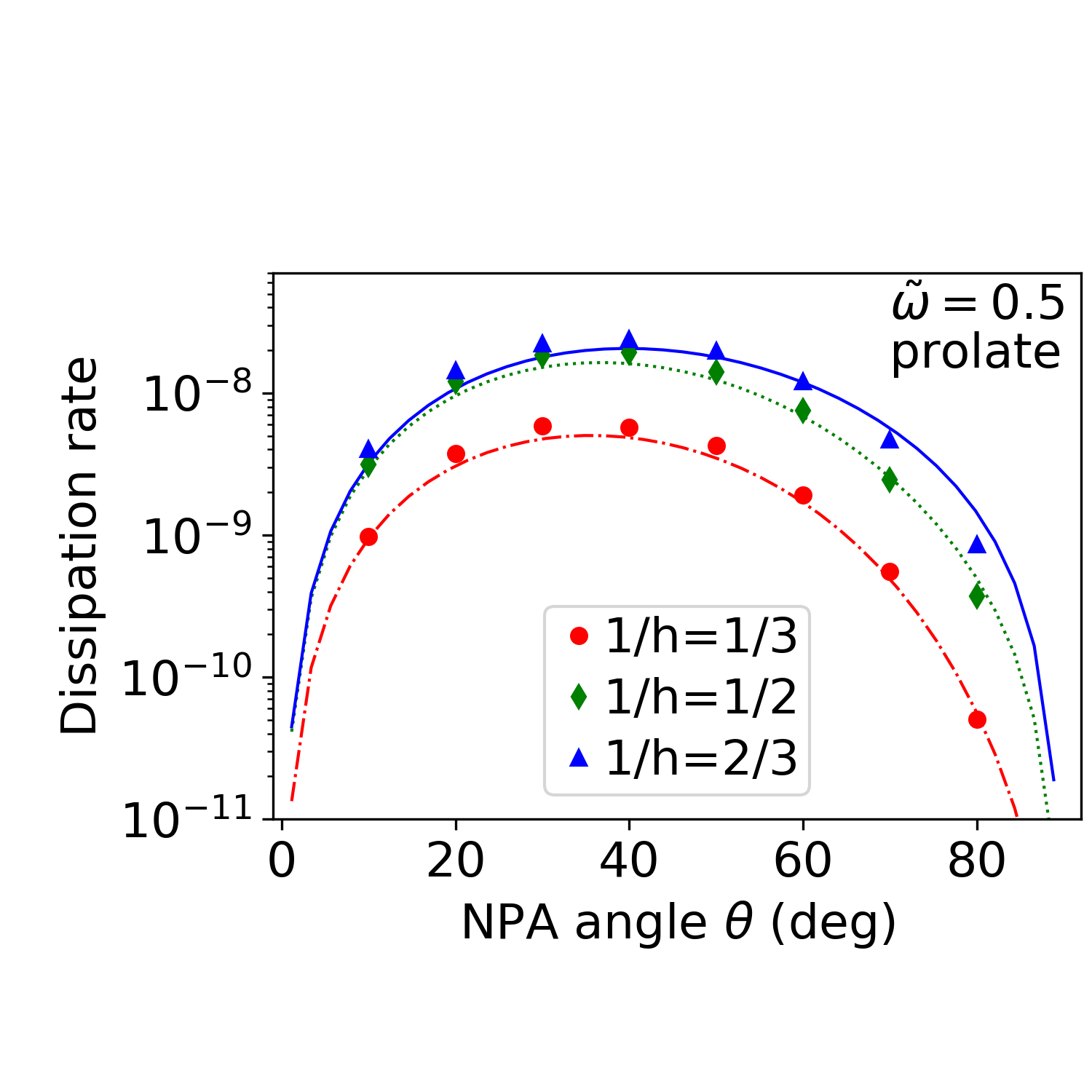}  
\includegraphics[width=3.25in, trim={6mm 10mm 0mm 30mm},clip]{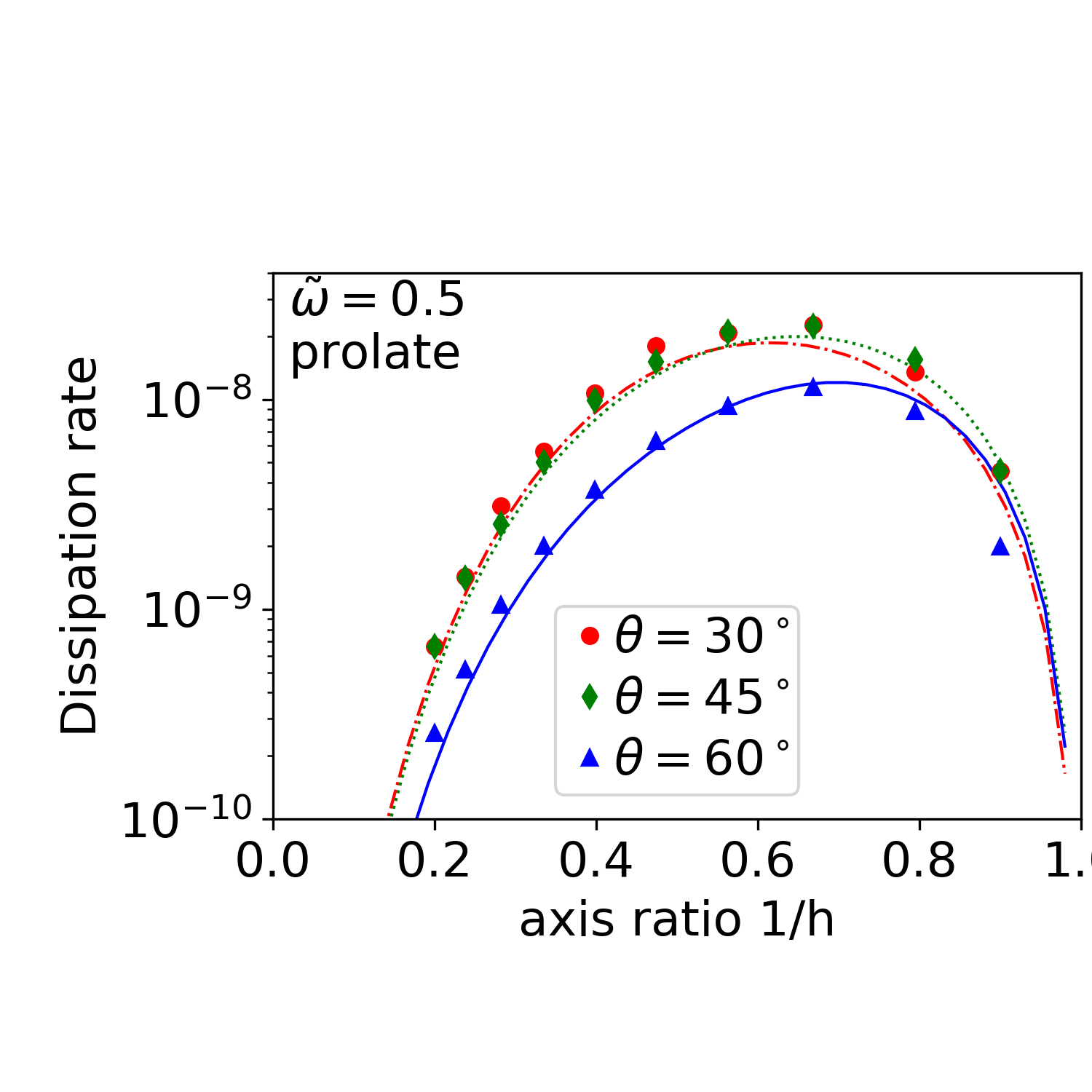}  
\caption{We compare the energy dissipation rates for prolate simulations that are identical 
except for their body axis ratio and NPA angle $\theta$.    Similar to Figure \ref{fig:oblate_BR} except
we show prolate simulations and the $x$ axis is the inverse of $h$.
The parameters for these simulations are listed in Tables \ref{tab:common} and Table \ref{tab:prolate}.
a)   The red circles, green  diamonds and blue triangles show measurements
for axis ratios $h=3,2, 3/2$, respectively.  
The red dash-dotted, green dotted and blue solid lines lines are computed for the same axis ratios 
 using equation \ref{eqn:BR3} based on work by \citet{breiter12}.
b) The red circles, green  diamonds and blue triangles show measurements
for NPA angle $\theta=30,45, 60^\circ$,  respectively.
The red dash-dotted, green dotted and blue solid lines lines are computed for the same NPA angles 
 using equation \ref{eqn:BR3}.
}
\label{fig:prolate_BR}
\end{figure}

Similar sets of simulations as described in section \ref{sec:oblate} were done for prolate ellipsoids instead
of oblates. Again we fix  frequency $\tilde \omega$, shear modulus and viscosity but vary
the body axis ratio and initial NPA angle $\theta$.
The simulations have
parameters listed in Table \ref{tab:prolate} and Table \ref{tab:common} and the energy dissipation rates for these simulations are shown in 
Figure \ref{fig:prolate_BR}. 
We ran 3 sets of prolate ellipsoid simulations with body axis ratios $h=3,2,3/2$ and 
 with NPA angles $\theta  = 10,20,30,40,50,60,70$ and $80^\circ$.
We also ran 3 sets  of prolates simulations with $\theta = 30, 45, 60^\circ$ 
and with 10 different values of axis ratio $h$.

Following equation 106 by \citet{breiter12} for prolates, 
equations \ref{eqn:BR1} and \ref{eqn:BR2} here can be modified to apply to prolates  by replacing $a$  with  $ c$,
and using our 
convection for prolate axis ratio $h>1$   (note that \citealt{breiter12} adopted $h<1$ for prolates).
For prolates equations \ref{eqn:BR1} and \ref{eqn:BR2}  should be proportional to $c^7$ rather than $a^7$.
For a prolate body with semi-axes $b=c$, volume equivalent radius $R_{\rm vol}$ and axis ratio $h=a/c$,  we find that
$c= h^{-\frac{1}{3}} R_{\rm vol}$.  Replacing $a^7$ with $c^7 = R_{\rm vol}^7 h^{-7/3} $ in equation \ref{eqn:BR2} 
gives equation \ref{eqn:BR3}, unchanged.  Equation \ref{eqn:BR3} should apply to both prolates and oblates, with
the convention that $h>1$ for prolates and $h<1$ for oblates.

On Figure \ref{fig:prolate_BR} we plot lines computed using equation \ref{eqn:BR3} (based on the
computations by \citealt{breiter12} but modified for the Kelvin-Voigt model) with
our numerical measurements, finding that they closely match our computations.

\citet{frouard18} used the stress tensor computed by \citet{sharma05}  for homogeneous ellipsoids
and so should be valid for both oblate and prolate 
ellipsoids.    
However, to apply to  prolates, the semi-major axis $a$ in equation 57 by \citealt{frouard18},  
equations \ref{eqn:P57}, \ref{eqn:PKV}, and \ref{eqn:PKV2}  should be replaced by
the semi-axis perpendicular to the axis of symmetry (replace $a^7$ by $c^7$). 
Replacing $a^7$ with $c^7 = R_{\rm vol}^7 h^{-7/3} $ in equation \ref{eqn:PKV2} 
gives equation \ref{eqn:PKV3_ob}, unchanged.  Equation \ref{eqn:PKV3_ob} should apply to  prolates and oblates.
Unfortunately equation \ref{eqn:PKV3_ob} does not match our numerical measurements, except at $h\sim 1$.
The curves are two orders of magnitude  higher than our numerical measurements at large $h$ and an order of magnitude higher at $\theta =45^\circ$.
We have not found errors in the computations by \citet{frouard18}, so we suspect that
 the stress tensor computed by \citet{sharma05} is a good approximation only in the limit of 
$h \la 1 $.   Both  \citet{sharma05} and \citet{breiter12} assume traction free boundary conditions but
their stress tensors differ (see the discussion in section 6 by \citealt{breiter12}). 

\begin{table}
\vbox to 115mm
{
\caption{\large Simulations of Inhomogeneous Rotators}
       \label{tab:core}
\begin{tabular}{llll} %
\hline
Fiducial model\\
\hline
Body shape &&  Bennu shape model \\
Spring constant & $k_s$ & 0.32 \\
Shear modulus & $\mu$  &  $ 5.2$ \\ 
Spring damping coefficients & $\gamma_s$ & 4 \\ 
Viscoelastic relaxation time & $t_{\rm relax}$ &  $ 3.8 \times 10^{-3}$ \\
Frequency & $\tilde \omega$ & 0.8 \\
NPA angle & $\theta$   & 50$^\circ$ \\
\hline
Variants &  & \\ 
Core properties &  Springs  & Viscoelastic parameters \\
\hline
Hard Core   &   $4 k_s$, $\gamma_s$   & $4\mu, \tau_{\rm relax}/4$ \\
Hard Core, similar $\tau_{\rm relax}$ &  $4k_s$, $4\gamma_s$ & $4\mu, \tau_{\rm relax}$ \\
Soft Core &   $k_s/4$,  $\gamma_s$ & $\mu/4, 4 \tau_{\rm relax}$ \\
Soft Core, similar $\tau_{\rm relax}$  &  $k_s/4$,  $\gamma_s/4$& $\mu/4,  \tau_{\rm relax}$ \\
Higher Viscosity Core & $k_s$, $4\gamma_s$ & $\mu, \tau_{\rm relax}/4$\\
Lower Viscosity Core & $k_s$, $\gamma_s/4$ & $\mu, 4\tau_{\rm relax}$\\ 
\hline
Relevant figure && Figure \ref{fig:core} \\
\hline
\end{tabular}\\
Additional simulation parameters are listed in Table \ref{tab:common}.
These simulations are discussed in section \ref{sec:core}.
The top section of the table lists parameters for a fiducial homogeneous model.  
We also ran similar simulations with harder and softer cores
and cores with higher and lower viscosity.   We changed spring constants or/and spring damping parameters only.
These variations listed in the second section of the table.    
\\
}
\end{table}

\subsection{Discussion on Material properties}
\label{sec:material}

As asteroid internal composition and behavior
is not well constrained, we take a moment to summarize laboratory estimates for shear modulus
$\mu$ and quality factor $Q$  in different materials and touch on estimates for $\mu Q$ in asteroids.
The wobble damping timescale is shorter for faster rotators, so more slowly rotating objects 
can persist in an NPA state for a longer time. 
The division between NPA
and principal axis (PA) rotators in the Asteroid Light Curve Database as a function of asteroid size 
and rotation period (see Figure 8 by \citealt{pravec14}) is used to estimate $\mu Q$ in asteroid material 
(building upon  \citealt{harris94}).   
Updated wobble damping timescales estimated by \citet{pravec14}
 from the division are 7-9 times shorter than those estimated
by \citet{harris94}, giving a somewhat lower rough estimate for $\mu Q \sim 10^{11}$ Pa, 
compared to the $5 \times 10^{11}$ Pa  estimated by \citet{harris94}.   
These estimates for $\mu Q$ can be substantially higher (by about two orders of magnitude)
than estimated from tidal dissipation
in binary asteroids (e.g., \citealt{taylor11,nimmo17}).  It might be difficult to resolve this quantitative disagreement
without better understanding of spin excitation, YORP, binary-YORP and binary formation and evolutionary processes.

To put things in context with stress/strains measured in laboratory experiments, we roughly estimate the sizes
 of stress and strain rates for asteroids undergoing NPA rotation, 
  \begin{align}
 \sigma &\sim \rho \Omega^2 R^2  \nonumber \\
 &=   5\ {\rm Pa}  \left( \frac{\rho}{1000\ {\rm kg~m}^3} \right)
\left( \frac{1\ {\rm day} }{P} \right)^2 \left( \frac{R}{ 1\ {\rm km}}\right)^2 \label{eqn:stress}
 \end{align}
 where $P = 2\pi/\Omega$.
 The  strain and strain rate
 \begin{align}
 \epsilon &\sim  \frac{\sigma}{\mu} \nonumber \\
 &   \sim 5 \times 10^{-9}   \left( \frac{\rho}{1000\ {\rm kg~m}^3} \right)
 \left(\frac{1\ {\rm day} }{P} \right)^2 \left( \frac{R}{ 1\ {\rm km}}\right)^2 \left( \frac{ 1\ {\rm GPa}}{\mu} \right)\nonumber \\
 \dot \epsilon &\sim \epsilon \Omega  \label{eqn:strain} \\
 & \sim 4 \times 10^{-13} {\rm s}^{-1} 
  \left( \frac{\rho}{1000\ {\rm kg~m}^3} \right)
 \left(\frac{1\ {\rm day} }{P} \right)^3 \left( \frac{R}{ 1\ {\rm km}}\right)^2 \left( \frac{ 1\ {\rm GPa}}{\mu} \right). \label{eqn:strain_rate}
 \end{align}
To be more accurate, the sensitivity to spin state and body axis ratios needs to be taken into account.
These values illustrate that oscillating stress, strain and strain rate during NPA rotation 
are quite low compared to laboratory experimental values that tend to have 
$\dot \epsilon \ga 10^{-12} {\rm s}^{-1}, \sigma \ga 10^{3}$ Pa 
(e.g., \citealt{goldsby01,mccarthy16}).

Mechanisms for dissipation
caused by constant stress and causing creep may differ from those caused by periodic stress/strain cycling
 (e.g., \citealt{cooper02,mccarthy16}).   Quality factor and shear modulus for different materials are likely
 dependent on amplitude as well as frequency.  The wobble damping regime is low stress and strain rate
and low frequency compared to seismic frequencies, and not necessarily similar to that for static stress.
Most laboratory studies find that $Q$ increases slowly as frequency increases, with a power law, 
$Q \propto f^{\alpha}$ and exponent $\alpha$ in the range 0.1--0.4 (e.g., \citealt{cooper02,mccarthy16}).
Equation \ref{eqn:Q} gives viscosity $\eta = \mu Q/\omega_{\rm prec}$, so a weak power law $Q$ dependency on frequency
partly cancels the frequency dependence of the viscosity, at a fixed shear modulus.
Using a spin period  of 10 days and its associated frequency, the viscosity $\eta \sim 1.4 \times 10^{16} $ Pa s
for $\mu Q = 10^{11} $ Pa.

For 'warm' ice at temperature $\sim 240 $ K,
laboratory experiments of low-frequency, small-strain periodic loading at
 Europa's orbital or tidal flexing frequency,  
$f \sim 2 \times 10^{-5}$ Hz, (corresponding to a rotation period of 3.55 days)
 give an  attenuation coefficient in the range  $Q \sim $  10 -- 100  \citep{mccarthy16}.
 The physical mechanism is chemical diffusion on grain boundaries or strain
accommodated intracrystalline dislocation slip. 
With a Young's modulus of a few GPa,  (and shear modulus about half this), the product $\mu Q$ for warm ice 
is similar to that estimated by \citet{harris94} and \citet{pravec14}.
Viscosity for colder ice would be a few orders of magnitude lower, so icy bodies in the outer solar system might take longer
to damp into principal axis rotation states.

However, ice fractions could be low for many asteroids.   
The Young's modulus for  solid rocky materials is about 50 GPa
and the $Q$ value at the same frequency regime for polycrystalline silicates may be similar to that of ice \citep{jackson02}.
The estimated $\mu Q\sim 10^{11} $ Pa would  require a lower $Q \sim 10$ 
value to be consistent with the NPA/PA division if asteroids are solid rock.

The Young's modulus of a porous rubble pile or a granular material can be an order of magnitude
lower than that of its grains  (see \url{http://www.geotechdata.info/}, \citealt{goldreich09}). 
Hydrostatic pressure, due to self-gravity, creates force chains 
through the medium, comprised of forces 
exerted at particle-particle contacts.  At low porosity, contacts can flex, and 
the granular medium can be softer than the individual particles themselves.  The medium need not
have tensile strength to exhibit viscoelastic relaxation when existing force chains are periodically varied at low amplitude.
Laboratory granular media can have high attenuation rates, $Q \sim 10$, for seismic waves (e.g., as does dry sand and silt;  \citealt{oeize02}),  though lunar regolith
has low seismic attenuation \citep{dainty74,toksoz74,nakamura76}, with $Q \sim 3000$.  
The lower the porosity and coarser the granular medium (having larger grains), the more it behaves like a solid.
The material strength could also depend on static hydrostatic pressure and consequently on the size of the asteroid.
If the Young's modulus is similar to that of ice or a few GPa, (and about a factor of 10 lower than
that of solid rocky materials) then $Q \sim 100$ would match the estimate 
for $\mu Q \sim 10^{11}$ Pa based on asteroid light curves.
The estimate for $\mu Q$  by \citet{pravec14}
lies easily within the large uncertainties if most asteroids are dry granular rocky materials.

The division between NPA and principal axis rotators in the Light Curve Database 
is better matched by a line of constant wobble damping time and does
not follow the collisional lifetime  or YORP time scale \citep{pravec14} so the excitation mechanism
has not been pinned down. 
\citet{pravec14} suggested that this trend could be explained if $\mu Q$ depends on asteroid size, 
decreasing  for smaller asteroids.
This can arise if larger asteroid cores contain lower porosity cores.  A larger
ice core fraction would make the core weaker but possibly more dissipative.
Stress, strain and strain rate (equations \ref{eqn:stress}, \ref{eqn:strain}, 
\ref{eqn:strain_rate}) all depend on asteroid size and all are larger for larger asteroids.  
If the dissipation rate is higher (giving a lower $Q$) at larger stress, strain and
strain rate, then $\mu Q$ could be lower for larger asteroids, opposite to the expected trend.

\citet{pravec14} adopted the quality factor ($Q$) based model 
for the wobble damping time \citep{sharma05,breiter12}  and giving  a damping time that is proportional
to the spin frequency to the minus third power, as in equation \ref{eqn:burns}; 
$\tau \propto \mu Q/ (\rho R^2 \tilde \omega^3)$ (also see equation 3 by \citealt{pravec14}).   Because
the energy dissipation rate for the Maxwell model depends on viscosity rater than $Q$, a constant
viscosity model gives lifetime for NPA rotation of $\tau \propto \eta / (\rho R^2 \tilde \omega^{2})$ \citep{frouard18}. 
The use of a constant viscosity $\eta$ rather than constant quality factor $Q$ model changes the relation between size
and spin frequency at a constant wobble damping lifetime and so would affect the slope of the NPA/PA division
on a plot of size versus spin frequency. 
At a constant wobble damping time
$R \propto \tilde \omega^{-\frac{3}{2}}$ for the constant $Q$ based estimates, but $R \propto \tilde \omega^{-1}$
for the constant viscosity and Maxwell based estimate.  
The Maxwell model by \citet{frouard18} slightly alleviates, but does not
remove altogether, the tension
between the distribution of NPA rotators as a function of size and spin rate, and that expected
if they all have the same material properties (and if NPA rotation is exited by collisions).

For ellipsoids with rotational symmetry, a Fourier expansion of the stress and strain contain only
two frequencies, the precession frequency and twice this frequency.  For triaxial bodies, the expansion
contains higher frequencies.  Since most materials show higher viscoelastic dissipation rates
at higher frequencies,  frequency sensitive viscoelastic wobble damping models would  
predict more rapid damping for elongated bodies than for more nearly spherical ones. 
Future and larger light curve databases might be used to reexamine the question of
how asteroid tumbling is excited and damped and how the dissipation in asteroid material
depends on frequency. 


The YORP effect acts adiabatically so would not
be expected to excite NPA rotation.  However, if a body is disrupted when spun up 
by YORP (e.g., \citealt{sanchez18})
then landslides and equatorial mass shedding might excite tumbling.
In this paper we have computed energy dissipation rates and have not integrated them to estimate
wobble damping times.  This integration is done assuming that angular momentum is conserved
\citep{sharma05,breiter12,frouard18}.  However YORP effect induced spin variations would violate
that conservation law.
There is a non-trivial connection between YORP, composition and rotation state (e.g., see \citealt{breiter15}).

\section{Inhomogeneous Rotators}
\label{sec:core}

In this section we measure the energy dissipation rate for nearly spherical wobbling bodies with hard and soft cores.
Inspired by the recent arrival of the OSIRIS-REx spacecraft at asteroid 101955 Bennu in Nov. 2018, we  
explore wobble damping using the radar based shape mode by \citet{nolan13}, as   
when studying  elastic response to impacts \citep{quillen19_bennu}.  We are interested in
the sensitivity of the wobble damping time to variations in internal composition.
The initial node distribution and spring network are created in the same way as for the ellipsoids, except we retain randomly generated particles
within the shape model; for more information see \citet{quillen19_bennu}.
This  shape  is nearly spherical, with moment of inertia ratios $\sqrt{I_1/I_3} \sim 0.94$ and $\sqrt{I_1/I_2} \sim 0.98$. 
Here $I_1<I_2<I_3$ are the three moments of inertia we measured numerically from our simulated mass node distribution.

We compared the energy dissipation rate for the Bennu shape model
with a triaxial ellipsoid model with similar moments of inertia and initial spin values, $\tilde \omega \sim 0.6$, and
NPA angle $\theta \sim 50^\circ$ (we computed the angle between the initial spin vector and angular momentum).
As the body is nearly oblate, the angle does not exhibit large variations during the simulation. 
Comparing a homogeneous Bennu shape model with the similar triaxial ellipsoid,
the numerically measured energy dissipation rates differed by less than 15\%.  
We conclude that shape variations for near spherical homogenous bodies 
do not strongly affect the energy dissipation rate during wobbling damping.

To probe the sensitivity of the wobble damping time to internal composition, we
varied the properties of core springs, leaving springs nearer the surface unchanged.
For these simulations we use the Bennu shape model, 
starting with a   fiducial and homogeneous model parameters with parameters listed in the top of Table \ref{tab:core}.    
We varied 20\%, 50\% or 80\% of the springs, changing their spring constants $k_s$, damping parameter $\gamma_s$
or both parameters.
We chose springs to change based on their midpoint positions.   For the 20\%,  50\%  and 80\% core models
we varied springs with midpoints within a radius of 0.60, 0.78, and 0.92, respectively,  
of the center of the body, giving harder or softer
or more less viscous cores compared to the fiducial model.  
The types of spring variations are listed in the bottom of Table \ref{tab:core}.
For hard and soft core simulations we
multiplied the fiducial model spring constant by 4 or 1/4.   For higher viscosity and lower viscosity
simulations  we multiplied the spring damping parameter by 4 or 1/4.  We also multiplied
both spring constant and damping parameter by the same factors to carry out simulations of hard and soft cores 
that have the same viscoelastic relaxation time as their shells.

\begin{figure*}
\includegraphics[width=7.25in, trim={30mm 10mm 0mm 30mm},clip]{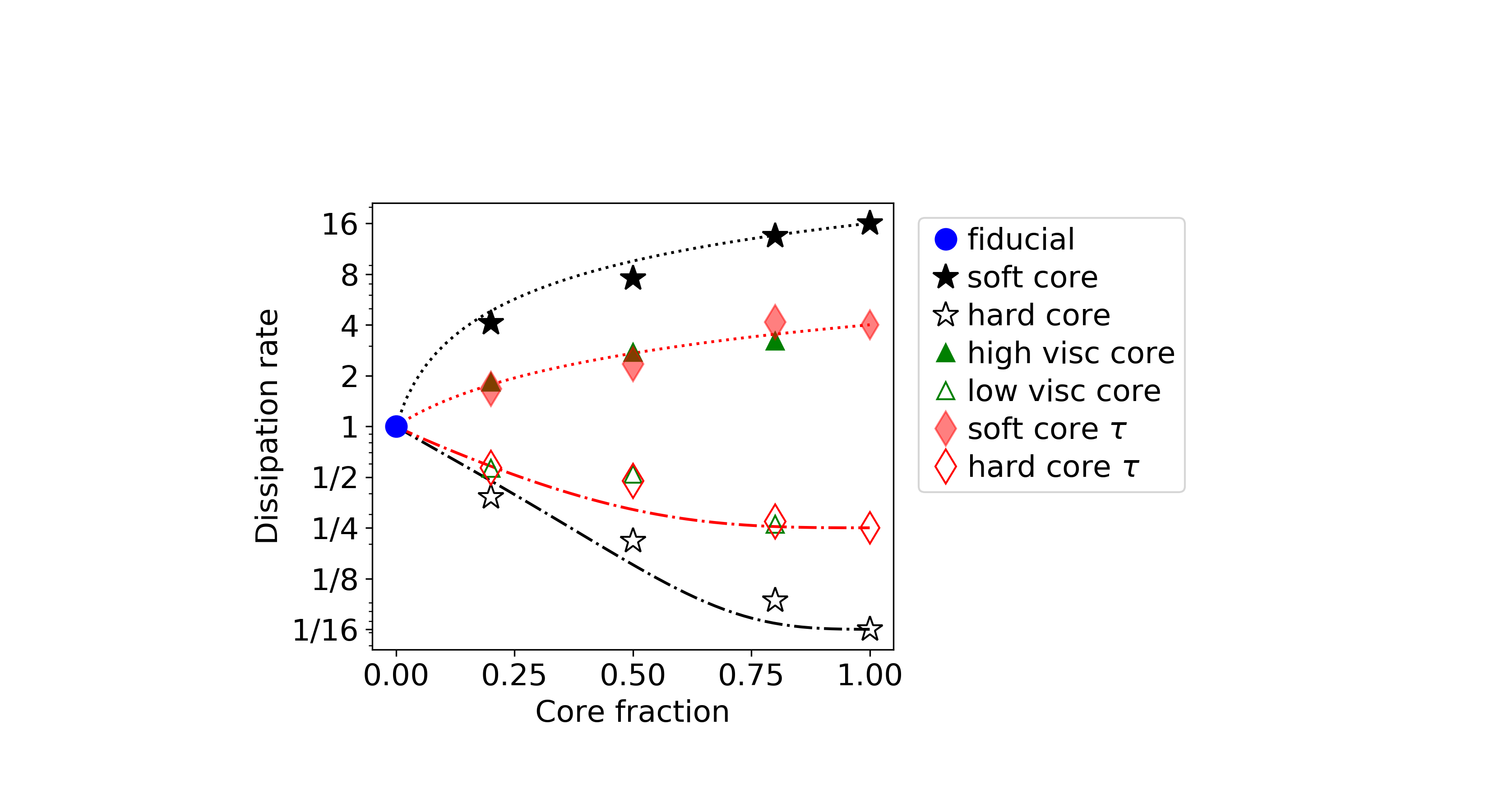}  
\caption{
Energy dissipation rates for simulations of core/shell inhomogeneous bodies based on the Bennu shape model.
The  $y$ axis shows the  the energy dissipation rates normalized to the homogeneous or fiducial simulation and with a log scale.
The $x$ axis gives the fraction of springs that are varied.  The homogenous fiducial simulation is shown as a blue
circle and has parameters listed in the top of Table \ref{tab:core}.   
Harder cores have solid or filled point types and softer cores have non-filled point types.
The types of cores are labelled in the key and refer to the descriptions in the bottom half of Table \ref{tab:core}. 
The dotted lines show z-models and the dot-dashed lines show r-models computed with equation \ref{eqn:fmodel}
that are based on quadratic models for internal stress.
}
\label{fig:core}
\end{figure*}

Energy dissipation rates for the hard and soft core models are shown in Figure \ref{fig:core}, normalized
to the dissipation rate measured for the homogeneous and fiducial model (shown as a large blue dot on the left). 
As the energy dissipation rate is proportional to $\tau_{\rm relax}/\mu$ (as in equation \ref{eqn:BR3}),
the homogeneous models with all  springs changed should have energy dissipation rates
4 or 16 times larger or smaller than the fiducial model.  Points on the right hand side of Figure \ref{fig:core}
are these factors of the homogeneous or fiducial model.
Figure \ref{fig:core} shows that  the energy dissipation rate from wobble damping is not linearly
dependent upon the core volume.  
The energy dissipation rate for wobble damping is more sensitive to the core material
properties than those in the shell. 
It is well known that viscoelastic tidal
heating is stronger at the base of a solid but dissipative shell (e.g., \citealt{beuthe13}), so we should not be surprised that
the energy dissipation rate for bodies undergoing NPA rotation is more strongly
influenced by material properties in the core than near the surface.

The stress and strain tensors computed by \citet{sharma05,breiter12} are quadratic in coordinates but
also include constant terms.   
The terms that are quadratic in coordinates drop to zero in the core
leaving the constant term setting the core stress and strain. 
We consider two simple models for the sensitivity of the energy dissipation rate to modified core properties.
We consider stress quadratically dependent on a single direction  (the z-model)
\begin{align}
 \sigma & \propto (1 - z^2) 
 \end{align}
 here aligned with the $z$ axis, 
 or quadratically dependent on the radius (the r-model)
 \begin{align}
 \sigma & \propto (1 - r^2) .
 \end{align}
Both models are traction free on the surface with $r=1$.
The actual stress tensors contain numerous quadratic terms (see the appendices by \citealt{sharma05,breiter12}).
We assume a energy dissipation rate per unit volume $p \propto \frac{ \tau_{\rm relax} }{\mu }  \sigma^2$  and integrate
over the volume of a sphere to find the total dissipation rate.  We assume a spherical boundary between core and
shell with a
core that has a viscoelastic relaxation time $\tau_{\rm relax,core}$ and shear modulus $\mu_{\rm core}$
and a shell that has  $\tau_{\rm relax,fid}$ and $\mu_{\rm fid}$.
The ratio of dissipation rates of the model with core compared to 
fiducial model (computed by integrating over volume) as a function of the fraction of volume in the core $f$
\begin{align}
\frac{P(f)}{P_{\rm fiducial}}    =  \left( \frac{ \tau_{\rm relax,core}} { \tau_{\rm relax,fid}} \frac{ \mu_{\rm fid}}{ \mu_{\rm core}} - 1 \right) 
  \frac{f\left(1 + a_2 f^\frac{2}{3} + a_4 f^\frac{4}{3} \right)} {\left(1 + a_2  + a_4\right)  } + 1, \label{eqn:fmodel}
\end{align} 
with coefficients 
\begin{align}
a_2 = - \frac{2}{5} & \qquad a_4 = \frac{3}{ 35} &  {\rm z- model}, \\
a_2 = - \frac{6}{5} & \qquad a_4 = \frac{3}{ 7} &  {\rm r- model} .
\end{align}

In Figure \ref{fig:core}
we have plotted equation \ref{eqn:fmodel} for the z-model with dotted lines for the  soft core
simulations.  These have $\tau_{\rm relax,core}/\mu_{\rm core} = $ 4 or 16 times that of the fiducial model. We plotted 
the r-model with dot dashed lines and with   $\tau_{\rm relax,core}/\mu_{\rm core} = $ 1/4 or 1/16
for the hard core models.   The z-model is a pretty good match to soft core simulations 
and the r-model is a pretty good match to hard core simulations.   
We lack a qualitative understanding on why a single quadratic stress model failed to cover both settings.
Until  analytical models are extended to cover core/shell models, these rough models can be used to estimate wobble damping times
in nearly spherical asteroids such as Bennu, but with core material properties 
differing from those near the surface. 




\section{Summary and Discussion}
\label{sec:sum}

In this paper we have carried out numerical simulations of homogenous oblate and prolate ellipsoids in NPA rotation states.
We have used a mass-spring N-body code, approximating a Kelvin-Voigt viscoelastic material,
to measure the rate of energy dissipation of ellipsoids undergoing NPA rotation.
By independently varying the spin rate, while keeping axis ratios, shear modulus, viscoelastic relaxation time and spin orientation fixed,
we measured how the energy dissipation rate depends on spin. 
We find that the dissipation rate is sensitive to spin to the 6-th power for the Kelvin-Voigt rheology 
rather than 4-th power as is true in the Maxwell rheology \citep{frouard18} or 5-th power as 
for the quality factor $Q$ based estimates \citep{breiter12}.
The sensitivity of the energy dissipation rate to rheology suggests that tumbling damping times are sensitive to the material properties of asteroids, and as a consequence the statistics of their spin states could be sensitive to their material properties, and as suggested by \citet{pravec14}.

By independently varying the spring damping rate, while keeping shear modulus, axis ratios, and initial spin vector fixed,
we found how the energy dissipation  rate depends on viscoelastic relaxation time and precession frequency.
The sensitivity of the dissipation rate to viscoelastic relaxation time (set by the spring damping)
allowed us to directly compare our dissipation rates to those predicted for a Maxwell rheology  \citep{frouard18}.

Our numerically measured values of energy dissipation for oblates range over more than 4 orders of magnitude
and  match the analytical predictions by \citet{frouard18}  as a function 
of NPA angle and axis ratio after they have been modified for the Kelvin-Voigt rheology.
They also match the quality factor based computations by \citet{breiter12} after modification.
The match is quite good even though bulk viscosity and gravitational compression are neglected in the analytical calculations
and both are present in the numerical simulations.
In our study of tidal spin down \citep{frouard16}, we suspect that our 
neglect of bulk viscosity in the analytical calculation gave a 30\% discrepancy
between predicted and measured spin down rates.   
Because the energy dissipation rate due to NPA rotation varies over many orders of
magnitude, a 30\% discrepancy would seem small on our plots.  
Our numerical measurements tend to be slightly higher than the
predicted values by \citet{frouard18} and this could be in part due to the neglect of bulk viscosity in the analytical model.   
In contrast, the predicted values by \citet{breiter12} give slightly higher dissipation rates than we measure numerically.  
The discrepancies between models and simulations are at the level of about 20\%, 
underscoring the remarkable accuracy of both analytical 
and numerical computations.
 
We carried out a similar series of simulations of prolate ellipsoids, finding that the $Q$ based model by \citet{breiter12} 
gives a good  match to our numerical measurements.   Our success in comparing numerical results
and theoretical results based on different linear rheologies motivated us to reexamine and discuss estimates
for the shear modulus and quality factor of asteroid material.  We primarily did this to highlight
the uncertainties in material properties of asteroids and because the recent analytical works \citep{breiter12,frouard18}
had not touched on recent laboratory measurements of $Q$ and elastic modulus.

In section \ref{sec:core} we carried out a short numerical exploration of the energy dissipation 
rate of  nearly spherical wobbling bodies with hard and soft cores.  We found that 
the energy dissipation rate for wobble damping is more sensitive to the core material
properties than those in the shell.  We found that an quadratic internal stress models
(see the appendices by \citealt{sharma05,breiter12}) matched the numerically measured energy dissipation rates
as a function of core volume,
though the form of the model differed between the hard core and soft core simulations.

We also explored (but do not discuss) long simulations of elongated but weak bodies 
following the progress of a single body as it relaxed
to principal axis rotation.   Occasionally we saw time periods with rapid
energy dissipation.  We attribute this to normal mode excitation, occurring when a spin frequency
matches a bending mode.  We have not focused on this phenomena here as it happens only in a soft body regime
where the normal mode frequencies are close to the precession frequency and asteroids tend to be
too stiff to be in this regime.
 However, we do want to bring it to the attention of
the reader as this capability of a mass-spring model code might prove useful in the future.  
Normal mode excitation would be necessarily ignored in most analytical computations of wobble damping and did not occur
during  the short simulations carried out in this paper, but can seen in our simulations when excited by an impact (e.g., \citealt{quillen19_bennu}).

In this study we have primarily restricted our simulations to ellipsoids with rotational symmetry.
For triaxial bodies, the acceleration in the body frame contains more than two frequency 
components.  This makes it more challenging to relate the numerically estimated energy dissipation rates
to  those of the Maxwell or $Q$ based wobble damping models.  This motivates future development of more sophisticated
numerical techniques to go beyond the mass spring model used here that approximates a Kelvin-Voigt  model.
Wobble damping in triaxial bodies is complicated by the transition between short axis and long axis rotation modes.
Even an infinitesimally small triaxiality from a prolate shape alters the relaxational dynamics, owing to the emergence of a
long period separatrix dividing the initial body's trajectories from the lowest energy end-state. Near the separatrix the tumbling-relaxation process slows considerably \citep{efroimsky01,efroimsky02}.
This makes future of study
of triaxial bodies undergoing NPA rotation particularly interesting.

\vskip 2 truein

Acknowledgements:

We thank Julien Frouard and Michael Efroimsky for helpful, constructive suggestions and comments that have corrected and improved this manuscript.
We thank  Eric Blackman, Esteban Wright, Larkin Martini and Randal Nelson for helpful discussions.

This material is based upon work supported in part supported by NASA grant 80NSSC17K0771,
and  National Science Foundation Grant No. PHY-1757062.

Code used in this paper is available at \url{ https://github.com/aquillen/wobble }    .

\bibliographystyle{mnras}
\bibliography{refs_oumou} %

\end{document}